\documentclass[sigconf,screen]{acmart}

\AtBeginDocument{%
  \providecommand\BibTeX{{%
    \normalfont B\kern-0.5em{\scshape i\kern-0.25em b}\kern-0.8em\TeX}}}

\copyrightyear{2024} 
\acmYear{2024} 
\setcopyright{acmlicensed}\acmConference[FAccT '24]{The 2024 ACM Conference on Fairness, Accountability, and Transparency}{June 3--6, 2024}{Rio de Janeiro, Brazil}
\acmBooktitle{The 2024 ACM Conference on Fairness, Accountability, and Transparency (FAccT '24), June 3--6, 2024, Rio de Janeiro, Brazil}
\acmDOI{10.1145/3630106.3658973}
\acmISBN{979-8-4007-0450-5/24/06}

\usepackage{hyperref}
\hypersetup{colorlinks,allcolors=black, breaklinks=true}

\begin{document}

\title[Epistemic Power in AI Ethics Labor: Legitimizing Located Complaints]{Epistemic Power in AI Ethics Labor: \\Legitimizing Located Complaints}

\author{David Gray Widder}
\email{david.g.widder@gmail.com}
\affiliation{%
  \institution{Digital Life Initiative, Cornell Tech}
  \city{New York City}
  \state{New York}
  \country{USA}
}

\renewcommand{\shortauthors}{David Gray Widder}

\begin{abstract}
What counts as legitimate AI ethics labor, and consequently, what are the epistemic terms on which AI ethics claims are rendered legitimate? Based on 75 interviews with technologists including researchers, developers, open source contributors, and activists, this paper explores the various epistemic bases from which AI ethics is discussed and practiced. In the context of outside attacks on AI ethics as an impediment to ``progress,'' I show how some AI ethics practices have reached toward authority from automation and quantification, and achieved some legitimacy as a result, while those based on richly embodied and situated lived experience have not. This paper draws together the work of feminist Anthropology and Science and Technology Studies scholars Diana Forsythe and Lucy Suchman with the works of postcolonial feminist theorist Sara Ahmed and Black feminist theorist Kristie Dotson to examine the implications of dominant AI ethics practices. 

By entrenching the epistemic power of quantification, dominant AI ethics practices---employing Model Cards and similar interventions---risk legitimizing AI ethics as a project in equal and opposite measure to which they marginalize lived experience as a legitimate part of the same project. In response, I propose\textit{ humble technical practices}: quantified or technical practices which specifically seek to make their epistemic limits clear in order to flatten hierarchies of epistemic power. 
\end{abstract}





\maketitle




\section{Introduction}

In a volume subtitled ``An Anthropologist in the World of Artificial Intelligence'' compiled after her untimely 1997 death, Diana Forsythe studies and critiques AI cultures in ways that still resonate today. She notes how the construction of what is considered ``knowledge'' in AI communities often ``deletes the social,'' privileging codified technical expertise over social knowledges and context~\cite{Forsythe_2001}, which in her analysis of the development of medical AI systems, led to their failure to usefully diagnose ailments. But more than two decades on, what were once ``toy'' systems in Forsythe's time are now the systems many are subjected to daily, even as both the definition~\cite{Suchman_2023} and functionality of AI remain questionable~\cite{Raji_Kumar_Horowitz_Selbst_2022}. In response, many companies and governments have AI ethics initiatives~\cite{Attard-Frost_De_los_Rios_Walters_2022,Jobin_Ienca_Vayena_2019}. This is not without conflict, as people leading these efforts have been fired: in perhaps the highest profile case, Timnit Gebru was fired from Google~\cite{Metz_Wakabayashi_2020} while attempting to publish a research paper raising concerns about the environmental and bias impacts of ever larger language models~\cite{Bender_Gebru_McMillan-Major_Shmitchell_2021}. Microsoft recently laid off one of its AI ethics teams, defending this decision by saying the central ethics team was broken up and individual workers moved ``within the individual product teams that are building the services and the software''~\cite{Schiffer_Newton_2023}. For those not fired, critical scholarship points to logics including techno-solutionism that constrain leading ethics programs at technology companies~\cite{Metcalf_Moss_boyd_2019}.

Integrating empirical interviews together with theory, this paper shows how the work of AI ethics is often understood as a ``subjective'' endeavor, and thus within the technology cultures in which this work sits, is cast as lower status when compared with the engineering work, seen as objective or rational.\footnote{The definitions of objectivity and subjectivity have taken nearly polar opposite meanings across history~\cite[p. 31]{Daston_Galison_2007}. The more familiar ``new'' senses of these words define so-called objective phenomena as not interpreted, and are thus taken to be  ``external to or independent of the mind''~\cite{Oxford_English_Dictionary_2023a}, and subjective phenomena are ``relating to the thinking subject ... proceeding from or taking place within the individual consciousness or perception; having its source in the mind''~\cite{Oxford_English_Dictionary_2023b}. In recognizing that all knowledge is situated and embodied, and thus not independent of the mind~\cite{Haraway_1988}, I reject this dichotomy even while recognizing it to be a social reality for many, including for many of my participants.} This is predicted by Forsythe's argument about the ``deletion of the social''~\cite[p. 28]{Forsythe_2001}, which she positioned as a detriment foremost to AI functionality, but in our context also operates to delete social knowledges from AI ethics discussions, and position such discussion as lower status. For women and other minoritized groups doing ethics work and disproportionately filling ethics roles, this interacts with documented ways in which their knowledge and work is devalued and delegitimized~\cite{West_Whittaker_Crawford_2019}. In response, I show how some participants seek to legitimize ethics work by casting it as objective, quantitative and therefore authoritative. However, I show how other participants attempt to speak about ethics from their own situated, embodied perspective, but find their attempts delegitimized. 

In some ways, scientific documentation of disparity in AI is crucial, and powerful: Buolamwini and Gebru's Gendershades project tested and documented gender and skin type disparities in facial recognition systems~\cite{Buolamwini_Gebru_2018}, which have now been cited in testimony before the U.S. Congress~\cite{Broussard_2019}, and in popular media outlets~\cite{Lohr_2018}, and is likely now one of the most widely recognized examples of AI ethics failings. Both Model Cards~\cite{Mitchell_Wu_Zaldivar_Barnes_Vasserman_Hutchinson_Spitzer_Raji_Gebru_2019} and Datasheets for Datasets~\cite{gebru2021datasheets} are widely seen as best practices, and have come to enjoy the authority of scientific ``Nutrition Facts'' on packaged food, both points recognized in practitioner blogs (``like nutrition labels [one can read a] model card, and quickly understand…'' ~\cite{Yi_2021}), and via pictographic representations shown on Google's prominent blog post about Model Cards~\cite{Google}.\footnote{ It is interesting to note that this Google blog post is based on a paper by Margaret Mitchell and Timnit Gebru (and others ~\cite{Mitchell_Wu_Zaldivar_Barnes_Vasserman_Hutchinson_Spitzer_Raji_Gebru_2019}), yet unsurprisingly, I can find no similar blog post bragging about a different paper which included these two authors~\cite{Bender_Gebru_McMillan-Major_Shmitchell_2021}, a paper at the center of their unceremonious firing from Google~\cite{Metz_Wakabayashi_2020}. More recent Google publications have failed to cite even the Model Cards paper, despite discussing Model Cards~\cite{Gemini_Team_2023}.  
} The examples above are extremely important to the extent to which ethics is taken seriously in technology cultures and organizations, and have become touchstones by which people communicate about AI ethics work. 

This push towards standardized processes can also be seen in attempts to codify AI ethics into rules and standards such as the European Union's AI Act~\cite{Madiega_2023}, and the US National Institute of Standards and Technology's ``AI Risk Management Framework''~\cite{NIST_2022}. This NIST framework asserts that while many characteristics that would make an AI system ``untrustworthy'' are ``directly connected to societal dynamics and human behavior,'' and that indeed, views on ``what makes an AI technology trustworthy differ,'' it nonetheless asserts that ``there are certain key characteristics of trustworthy systems,'' such as ``fair … accountable and transparent''~\cite{NIST_2022}. In this, the NIST framework provides for the contextual nature of trust~\cite{widder2021trust} to some degree, but then narrowly circumscribes this in a way perhaps more insidious than its wholesale omission. It acknowledges situatedness in how different people may understand what trust is, only to then reduce trust shallow techno-solutionistic questions of design~\cite{Greene_Hoffmann_Stark_2019}, by asserting ``fair ... accountable and transparent'' as universally important principles, with a stable and shared meaning, even while other research shows how the political economic context of inter-firm AI development shreds this minimal context~\cite{Gansky_McDonald_2022}.

However, if we conceive of ethics as an inherently and productively contested domain, as past work has called for~\cite{Young_Katell_Krafft_2022}, this push for legitimacy through codification and quantitative objectivity is likely to be counterproductive and harmful. 
This is especially pernicious for members of minoritized groups seeking to raise ethical concerns grounded in their lived experience, yet who are not afforded a ``view from nowhere,'' an epistemic privilege Timnit Gebru discusses in an AI context in her 2021 lecture~\cite{Gebru_2021}.
This will also serve to underwrite exclusionary pedagogy by delegitimizing ways of knowing from disciplines beyond computer science, as Raji and coauthors demonstrate in their analysis of AI ethics course syllabi~\cite{raji2021you}.
To begin to interrogate these politics of objectivity and examine epistemic power in AI ethics labor, 
the theoretical orientation of this paper begins with Lucy Suchman's notion of Located Accountability~\cite{Suchman_2002}, which draws on Donna Haraways' Situated Knowledges~\cite{Haraway_1988} to envision a feminist ethic of accountability in technology production enabled by locating oneself in this process, and remaining engaged with–and prepared to intervene in–the effects of this technology. Further, insofar as AI Ethics is cast as, and indeed can often become, a complaint voiced in opposition to existing development trajectories or business imperatives~\cite{Widder_Zhen_Dabbish_Herbsleb_2023}, raising ethical issues may be seen as a complaint. For this, even while she writes about harassment complaints in universities rather than technology ethics, I look towards postcolonial feminist theorist Sara Ahmed's study of complaint activism~\cite{Ahmed_2021}. Ahmed shows how without a concerted effort towards ``reorientation,'' complaints must be expressed in ways legible to institutional logics, in turn further entrenching the epistemological supremacy of the logics that had refused to take these complaints seriously. I explore ways to engage in this reorientation in the later sections of this piece, by hearing intersectional feminist Audre Lorde's call to not to reach for sources of legitimacy that reinscribe oppressive systems~\cite{Lorde_1984}, and Black feminist theorist Kristie Dotson's understanding of the intractability of flattening epistemic power differentials and suggestions for ways forward~\cite{Dotson_2014}. 

The empirical basis for this paper is 75 interviews with technologists conducted remotely between early 2021 and late 2023, in a multi-sited approach~\cite{williams2014multisited}. Purposive and snowball recruitment~\cite{palinkas2015purposeful} helped reach participants in a variety of contexts: 
the majority work in companies ranging from startups to multinationals, many work on open source projects, several are academics, and a smaller group are activists seeking to question local development of carceral technology. Many inhabited and moved between multiple such categories.
Participants were in four continents: the majority based in North America and Europe but also including those in Asia and Africa.
The majority were doing AI work at some point in its ``supply chain''~\cite{cobbe2023understandinga, Widder_Nafus_2023}: spanning dataset curation, academic AI research, building new models, to deploying them in consumer or B2B contexts. Corporate participants were often required to engage with corporate AI ethics practices. This includes some who have accepted AI ethics as part of their work, whether through intrinsic motivation or as part of their assigned duties~\cite[e. g.,][]{Metcalf_Moss_boyd_2019}, and many who have not. Interview questions were tailored to context and did not adopt consistent predetermined structure\cite{seidman2006interviewing}: some portion of participants spoke to me about their own ethical concerns, often after raising them to their employers and seeing them go unaddressed.  This study received IRB or other ethics approval as appropriate.

The variety of contexts in which I investigate AI ethics practices allows us to understand how a variety of epistemological stances interact with the legitimacy of AI ethics claims, and what is rendered possible or not as a result of these stances. Participants were invited to self-identify their gender: 57 identified as men, 13 identified as women, four identified as non-binary, and one declined to answer. Most interviews lasted one hour (though ranging between 25 minutes and 3 hours), and were transcribed and analyzed using inductive thematic analysis~\cite{Braun_Clarke_2006,Braun_Clarke_2019}\footnote{In an \textit{inductive} thematic analysis, themes are grounded in and emerge from the data, as opposed to \textit{deductive} analysis where themes are decided \textit{a priori}, and is properly a \textit{reflexive} process, where the researcher reflects on their assumptions, training, values, and how this may shape analysis. See also:~\cite{braun2021one, Braun_Clarke_2006}.}, under an \textit{interpretivist} epistemological paradigm~\cite{Lincoln_Guba_2000}.\footnote{
Interpretivism, similar to its younger cousin constructivism~\cite{schwandt1994constructivist}, understands that individuals are experts in interpreting their own experience, asks researchers to deeply understand these interpretations, and with both sharing the task of constructing wider meaning from these. To positivists, this is an unacceptable limitation, as no view from nowhere objectivity is possible. See also~\cite{Lincoln_Guba_2000, Daston_Galison_2007, Gebru_2021, Haraway_1988}.}
My analysis is therefore inflected by my own positionality: I am a male-identified scholar in the United States with formal training in computer science, mentored by feminist STS scholars, with experience organizing against pervasive surveillance~\cite{guo2023computer, gulland2023campus} and carceral technology~\cite{davidson2020pittsburgh,kubit2022cmu}. It is from attempts to reconcile differences in language and legitimacy between the first and last of these–computer science and carceral tech activism–that I am motivated to write this paper, but my positionality necessarily limits those I can talk with and inflects the interpretations I arrive at, and others may arrive differently.  
In some ways, these locations represent the ``stepping in and stepping out'' characteristic of fieldwork~\cite{Powdermaker_1966} in that some of my nonbinary and woman-identified participants construct their AI ethics practice as self-consciously feminist or even necessarily constructed in femme-only spaces, yet also, the predominantly technical and male backdrop in which AI ethics work sits paints me as part of the ingroup relative to the bulk of participants in this work: men in technical roles. 

This paper proceeds as follows: I first show how AI ethics labor is often seen as a chore, in contrast to the ``real work'' of building AI systems. I then show how attempts to legitimize this work by casting it as a quantitative, authoritative and even an automated project have succeeded to some extent in ``elevating'' it, by constructing it in the same terms ``real'' AI engineering work is constructed. I then turn to attempts by some participants to make ethics complaints in ways that eschew quantification and instead come from the insight of their own particular location and experience, and recount how such efforts are often delegitimized. Then, I examine how some participants seek to construct alternative AI ethics practices, often based on their embodied experiences, and largely but not entirely beyond objective institutional logics of legitimacy. I conclude by motivating and explaining humble technical practices as those which specifically seek to make their epistemic limits clear, sketching ways they may be implemented, and placing such practices in dialogue with work on liberatory uses of quantification, critical technical practices, participatory and intersectional AI, and epistemic power relations in AI. 

This paper contributes the following: 
\begin{itemize}
    \item      This work provides an empirically-based catalog of AI ethics labor conducted from a variety of epistemological stances and contexts, and examines their sources of---and the extent to each is afforded---legitimacy, complementing scholarship which scrutinizes hierarchies of knowledge in AI ethics educational contexts~\cite{raji2021you},  the supremacy of quantification in AI fairness work in corporate contexts~\cite{deng2023investigating}, and the use of seemingly ``objective'' algorithms in public sector~\cite{pruss2023ghosting}.
    \item Flowing from this, this work attempts to frame discussion of hierarchies of knowledge between the wide range of academic disciplines (ostensibly) included in AI ethics discourse: from those concerned with ``the development and deployment of algorithms'' to those ``examining human experiences, needs, perceptions''~\cite{acm}, and between these and attempts to include knowledge from activist and lived experiences outside the academy~\cite{acmb}, by examining how these different knowledges are (not) afforded legitimacy in practice.
\item Excellent exceptions notwithstanding~\cite[e.g.,][]{Suresh2022,Kong_2022,amugongo2023invigorating, sampson2023representation, png2022tensions, hancox-li2021epistemica}, reflexive analysis of AI ethics scholarship argues that more space needs to be made for feminist, Black, decolonial and non-Western thought ~\cite{laufer2022four,birhane2022forgottena,raji2021you}.  In its use of theory to interpret empirical data, this paper demonstrates the relevance of feminist STS~\cite{Forsythe_2001, Suchman_2002, Haraway_1988}, post-colonial scholarship~\cite{Ahmed_2021, Verran_2013}, and Black feminist thought~\cite{Dotson_2014, Lorde_1984} to AI ethics questions. 
    \item   Using this theory, this paper demonstrates the role of \textit{epistemic power} in the context of AI ethics labor,  joining work bringing an epistemic lens to algorithmic fairness~\cite{edenberg2023epistemic}. This responds to reflexive examinations of past AI ethics scholarship calling for greater attention to structural power asymmetries~\cite{birhane2022forgottena,laufer2022four} and growing examples of research on other kinds of power in AI ethics, such as in algorithmic decision-making~\cite{kasy2021fairness}, workers raising ethical concerns~~\cite{Widder_Zhen_Dabbish_Herbsleb_2023} and engaging in collective action~\cite{boag2022techa}, corporate capture co-opting academic critique~\cite{Young_Katell_Krafft_2022}, and between those subjected to algorithms and those developing them~\cite{benjamin2022fuckthealgorithm}. This localization of epistemic power in AI ethics labor is used to motivate and sketch the concept of \textit{humble technical practices}, drawing on related past work on critical technical practices~\cite{Agre_1997,Malik_Malik_2021}. 
    \item Finally, in examining epistemic power hierarchies, this paper provides a deeper diagnosis for the move to techno-solutionsistic practices of AI ethics, widely critiqued as problematic and insufficient~\cite{Greene_Hoffmann_Stark_2019, abdurahman2021calculating, Metcalf_Moss_boyd_2019, mittelstadt2019principlesc,laufer2022four}, by locating one reason for this move at the level of epistemic power.
\end{itemize}

\section{The lower status of ethics work within AI cultures}

This section demonstrates how my participants understand ethics work to be lower status relative to ``real'' AI work, but similar phenomena have been examined before. 
Past research shows how certain work in technology communities, like writing documentation or building community, is often done by women, and is less valued than writing code~\cite{Nafus_2012}. More recent work within AI has even shown differences between high status work on models versus lower status work on the dataset models are built from~\cite{Sambasivan_Kapania_Highfill_Akrong_Paritosh_Aroyo_2021}, and recent manifestos have attacked ``tech ethics'' on the basis that it hinders ``progress'' on endeavors such as ``Becoming Technological Supermen''~\cite{Andreessen_2023}. Forsythe shows how these status divides are inscribed in the language used by the AI practitioners she studied: they restricted their use of the word ```work' to what they think of as ‘real AI': building systems'' ~\cite[p. 26]{Forsythe_2001}, set in contrast to the other kinds of work such as grant writing, meetings, and also the ``knowledge elicitation'' interviews seen as boring but necessary to ``capture'' knowledge to encode into systems. 

When asked about whether his community of deepfake content creators had ever discussed ethics, one participant replied: 
``Uh, no. No. I don't think so. […] It's mostly technical, you know, stuff and, and sharing information about models. Um, another metaphor I use is like, it's like long exposure photography, but like from like the 19th century where you're carrying around plates, you know, and you have to be like a pristine, like technician in handling all the parameters to set up your camera and everything.''
In this metaphor, he explained how the work deemed worthy of discussion is focused on the technical craft of making realistic deepfakes, not the ethical questions involved in doing so. Notably, these two are seen as distinct and separate, a stark example of the more widespread division between technology and society~\cite{Latour_1993}. This community was overwhelmingly men, in contrast to other communities examining the gendered impact of deepfakes~\cite{Maddocks_2020}.

I often asked participants what work they were most excited about, and what work they saw as a chore. The most common task seen as a chore was ``paperwork'' needed to release new systems, which notably included AI ethics questionnaires, often included filling in model cards. One described how each of the steps needed to release code could be a Byzantine ``trap,'' and that each task was meted out randomly to teammates as necessary and undifferentiated homework. Another referred to this as ``light work,'' and another said that this work did not ``feel like I was doing something that I need to do,'' but merely that it was a ``have to'' do. 

One woman machine learning engineer described how she had once previously gone through these release steps, and now she was the one pigeon holed into doing these tasks evermore: ``It's mainly me on my team who's going through the [release steps…] I had to figure it out for one product and after that it has always been, [<name>] you know how to do it. You go ahead and do it.'' Forsythe wrote about how women and their work in AI labs were marginalized: they were administrators or secretaries, and by excluding their work from the category of the ``real work'' of building AI systems, they were not included as members of ``the lab.'' In one case, she recounts how they were ``rendered invisible through the systematic bracketing out of people who perform work that is gendered as female''~\cite[p. 172]{Forsythe_2001}. While our previous participant had a masters degree and an AI engineering job title, she was nonetheless the only woman on her team and the one told to do the work coded as a ``chore.'' Taken together, these examples show how AI ethics work is seen as a lower status ``chore.''

\section{Automated model cards: legitimacy via quantified objectivity}

Whereas in the last section, we saw how doing ethics work is often accepted as an unwelcome but necessary chore, here we see how others seek ways to legitimize its value. In her study of sexual harassment in universities, Sara Ahmed wrote how students' complaints were perceived as ``legible'' to their institution only when expressed in a certain way, using ``evidence that took the right form''~\cite[p. 5]{Ahmed_2021}, and often faced threats to their career when raising these complaints. In an AI ethics context, Madaio and coauthors found that advocating for AI fairness issues can bring perceived career risks, but that checklists can empower individuals to raise these issues~\cite{Madaio_Stark_Wortman_Vaughan_Wallach_2020}.

For some, pushing for legitimacy for AI ethics involved seeking to make ethical evaluation into an ``objective'' process through the use of automated tools. One AI engineer, on reviewing his organization's model card template, which already sought to walk through the kinds of questions which should be asked when thinking about ethics, lamented that ``A lot of filling out the model card is subjective, right, because it's based on my understanding of how we use the model and my understanding of the data, the data set.'' When I asked why this might be an issue, he said ``because, if I give the card and the data set and the model to another person to answer the question, his answer might be different.'' To him, ethics ought to be decidable and rational, and therefore he believed model cards ought to be able to conclusively ``answer'' questions of ethics, rather than open up such questions for contestation.\footnote{ While the original paper proposing Model Cards ``specifically aims to \textit{standardize} ethical practice and reporting'' (emphasis mine), the paper does so in order to enable questions to be asked  ``across different institutions, contexts, and stakeholders,'' and explicitly positions it as one technique among many needed interventions~\cite{Mitchell_Wu_Zaldivar_Barnes_Vasserman_Hutchinson_Spitzer_Raji_Gebru_2019}.} To help achieve these ends, he said how on his team ``we try to automate a lot of that now'' by building ``tooling to run the data and the model [...] that's actually, objectively looking at it,'' calculating numbers to generate a model card. Later in the interview, he reflected ``I haven't really figured out if you actually need a person to answer these questions, or the tooling can answer these questions for you.'' This was also apparent in how he spoke of ethical issues, for example talking of ethics issues as if they could be straightforwardly fixed as soon as better tools were available to enable model explainability, saying that ``The de-bug-ability of it so that at least you can fix the problem'' is most pressing to him. Finally, when asked what came to mind when he heard ``ethical AI,'' his first response was ``fairness,'' saying that it is ``based on a lot of definitions that we have around nondiscrimination … policy'' , attending to definitions and policy as first order concerns. 

Automating model cards was one way to cast ethics as quantitative and objective, but others sought to base their claims to AI ethics by citing evidence from outside authorities. In one group discussion about AI ethics I witnessed in particular, participants persistently backed up points they raised by citing a variety of outside sources, including: the book Thinking Fast and Slow, the book Bowling Alone, the notion of the Dunbar Number, and ``a study done way back in 1990s, 1998 or 96, something like that.'' In a shorter follow up conversation one-on-one, a participant brought up ``Checklist Manifesto. Have you read that book? It's a fantastic book. It's by Atul Gawande,'' and ``this website called [the] AI incidents'' database. In this way, ethical arguments were seen as especially valid when they were based on an external, and often scholarly, reference point. The reference to the AI Incidents Database~\cite{McGregor_2021} is particularly interesting: it is a resource which helps to expand the frame of discussion by providing evidence for the contextuality of AI harm, but its invocation suggests the need for such evidence to be particularly structured and collated to be seen as especially legitimate. 

Finally, the participant in a professional community committed to making ``quality'' deepfakes (discussed above), when asked about ethical issues, said that he is ``bothered'' that most discussions of ethics of deepfakes center on geopolitical misinformation concerns (eg, faking politicians) rather than the predominant case of deepfakes: non consensual porn~\cite{ajder2019state} (this discrepancy examined using a feminist lens in prior work~\cite{Maddocks_2020}). To him, ``objective'' debate was needed: ``like let's have an objective, you know, conversation about the benefits and the disadvantages [of deepfakes…].'' At the same time, he reasoned that most deepfake porn may not be high quality enough to nor common enough to be concerning in his desired ``objective'' debate: ``I know it's like 90\% is like pornography, but it's also, you can have a debate about, well, how many users are doing that? No one really knows, you know? And, and to be honest, those fakes aren't [... at] the level that the [professional] people that I work with.''

\section{Grey skin as technofix: failures to lodge located complaints}

It is analytically notable that both Ahmed and Suchman use the word ``location''' Suchman calls us to acknowledge the located nature of our ``vision''–that our perspective on the world is not a ``view from nowhere'' but is instead based in our particular social, political, geographical, and embodied worlds~\cite{Suchman_2002}. Ahmed speaks of location to denote the misdirection of blame: ``to become a complainer is to become the location of a problem''~\cite[p. 3]{Ahmed_2021}, in how a complainant might be told that they are the problem for misinterpreting things or raising a fuss, instead of treating the harassment or harasser as the problem. Both Suchman and Ahmed use ``location'' to denote the particular partial perspective from which one's views are generated, in Ahmed's case denoting the particular cues by which a student is attuned to a pattern of harassment mistakenly read as innocent to the ``objective'' admin who would adjudicate the matter. Those ``too close'' to the harassment were seen as unable to be objective.

Many participants report that when raising ethics complaints that are based on their own particular location, these are often delegitimized. For example, a woman of color working on virtual reality tech (the only woman of color on her team) asked for small steps to make their tech more inclusive for people of different skin colors, or those missing a hand. She reports: ``And virtual reality by itself, I will say is not very accessible, right? [...] So a really famous VR software at the time, had done inclusivity, in terms of the color of the skins of hands, right, and allowing for people with one hand to operate it. So if you didn't want to use white hands or stuff like that, and I brought it up as an option, because we were working with folks who could be helped by that.''

In line with the previous section, she had first tried citing ``papers on why diversity helps on how people will come back to your application,'' attempting to legitimize her concern using both outside scholarly authority and the valued business imperative of attracting more users. The work to legitimize this complaint by using the kind of evidence and imperatives seen as legitimate by her company constituted labor, which fell to her. But this approach failed, so this participant noted to her team how this concern was rooted in her own usage of the system, seeing white hands projected in VR in place of her color: ``I don't have white skin and the device did and I was like: Wait, if I'm developing it, if people who are using it, [they] might feel even more intensely about it.'' On one hand, she noted how ``it was easier to sort of bring up those diverse opinions'' by virtue of her identity, but on the other, she noted how she was concerned that others would ask whether her ``perspective [is] coming because I am a unique person [...] or is it only coming because I'm a woman and and then a woman of color?'' As a result, she said that she didn't want to ``rock the boat too much'' for fear these rocks would be dismissed by virtue of being ``quirks'' of her identity. Indeed, she said ``I think [that] sometimes I could be considered intense.''

Ultimately, the accessible and inclusive VR features she asked for were not implemented, ``And then I was sort of told, Well, nobody asked for it.'' Here, ``somebody'' would refer exclusively to paying users, meaning she was included in the empty group of ``nobody.'' Instead of customizable skin colors, the skin was made grey, which she named as a ``colorblind'' outcome, and disliked. This sidestepped situated discussions of skin color, and instead, adopted grey skin as a kind of objective skin from literal nowhere. 

Ahmed writes that ``we realized that to be heard we had to make our experiences legible [...] For some of us, the institution expected emotion and hurt to be expressed. For others [...] emotional expressions were viewed as irrelevant or even detrimental to complaint.''~\cite[p. 267-268]{Ahmed_2021}. While the context of sexual harassment in universities is different than the context of this participant, she was caught in a similar double-bind in choosing how to base her claim: on one hand, she had tried to cite academic evidence on why inclusion would help product success. On the other hand, she had attempted to speak from her own situated perspective, as a woman of color, for why she did not want VR to white or grey-wash her hands, but instead was implicitly called ``nobody'' — outside of the role of the customer, her own ask for more inclusive features did not count.

\section{Alternative AI ethics: space for embodied complaints}

Upon thinking back to an AI ethics conversation she had convened with exclusively femme work colleagues, one participant reflected how this changed the tenor of discussion, creating a shared ``feeling like it's [...] a part of your lived experience to be harmed by technologies, versus feeling like it's an abstract intellectual conversation.'' She went on to say how shared experience may make people feel comfortable to ``give more of yourself,'' because she can ``feel more comfortable when you're sort of assuming that the people in the room that are there with you are on the same page'' because of shared experience of marginalization, even while particulars may differ. In this way, she felt that discussing AI ethics with other marginalized people legitimizes speaking from ``lived experience'' instead of conceiving it as an ``abstract intellectual'' conversation. Here, gaps left by the latter are filled by the former, and it is important to note that even those engaging in ``abstract intellectual'' conversations do so from lived experience, but are simply more limited in what they recognize that this experience teaches them. In a recent discussion, she raised ``foreign intervention'' as a concern for discussion, based in part on the fact that ``I'm from Latin America, my family's from Latin America originally. And I think a lot about the impacts of technology on a global scale.'' She contrasted this recent discussion with past experiences of group discussions at her workplace which included men: ``it's also worthwhile to like, point out that we're all women. [...] there's a different way these conversations happen [...] when there's like, male genders participating in activities like this. [...] men take up space in particular ways.'' 

One participant, who identified as a ``grumpy feminist anthropologist'' told of her past experience taking an AI ethics class, where student groups would sign up to present and lead discussions of tech ethics issues in the news. She mentioned how she ``was paying attention to [the] gender component of those discussions.'' Noting that there were ``much more men than women'' in the class generally, and that in particular: 
``the people who signed up for discussion on [the article] ‘Google will know I'm pregnant before I [do]' were men. And they were talking about periods and being pregnant and all these things without like, any personal experience.''
She discussed how space was not created to discuss this topic from particular personal experience, instead treating all discussants as having equal---detached and supposedly objective---perspectives on the issue. She goes on: ``There was no acknowledgment of [the] embodiment of the experience when it comes to ethics and AI,'' mentioning that this was ``pretty memorable [...] as a woman and as somebody who gave birth and was pregnant.'' Speaking of a more recent AI ethics conversation, she suggested the way it was was constructed limited discussion in 
``what kind of terrain, what kind of habits, what kind of people'' fit and did not, giving an example of a local street vendor selling juice at what used to be a rich and busy gathering spot, but whose livelihood has been taken through the uptake of delivery apps. Speaking how the ``example of [the] juice maker doesn't fit in here,'' she characterized the way the conversation was constructed as an ``epistemic kind of violence'' in how it adopted and tacitly enforced a ``certain imagery of the culture.'' A different participant spoke similarly, reflecting on how she often put forward her experience as an immigrant when discussing technology: ``Sometimes I do it very consciously to like, put my foot in the room or like to remind [of] this context, that it affects everything. Like, everything that happens and gets built in the US is going to get exported as an idea if not an actual product into the rest of the world and into the global south.'' She spoke of how she does this even when conspicuously ``conscious about how I, in this country, I'm talking about this ‘other place' like I'm often the only person bringing up an ‘other place','' aware of how this may ‘other' her but seeing this as important nonetheless.

Others spoke concretely about how embodiment and situated experience played a role in their own development process, and how this intersected with corporate logics. One participant was developing a body scanning health product, mentioning: ``body scanning in particular is sensitive information, and so we recognize from the beginning that if you show somebody a body scan, number one, different people are going to react to it differently. There is great potential to have someone get motivated or track progress. However, there's also great potential to exacerbate body image issues. So that's something we're very sensitive to in the way we designed the app and the way we present data.'' He described the ways his team had sought to design the interface to account for these differing reactions, including a ``multi screen tutorial that walks the user through; how should you think about this information? You might be shocked by this information. That's OK. This is about you achieving your goals. And for each piece, each measurement, for example, we would provide in that tutorial, an explanation. Here's your waist circumference, here's why it's important, and here's a bit of research around it.'' While echoing discourses of legitimacy through scholarly medicine as seen above, this participant spoke of how how as part of the ``venture incubation process'' they would themselves volunteer to serve as each other's pilot participants, recognizing how much of it is ``a personal experience having [one's body] evaluated [by] 3D body scanning in a research context.'' He spoke of how they had developed mutual vulnerability through ``the opportunity to get to know each other. And so honestly, we have a pretty good team rapport. [...] we check in on each other outside of work, and so having that familiarity is definitely helpful, in talking about this.'' However, he also spoke about how a shared ``passion for the user, for the customer'' motivated the attention to the myriad ways people may relate or react to this technology, interestingly, suggesting attention to various experiences can be cast as a market imperative.

\section{Conclusions: towards \textit{humble} technical practices}

``The master's tools will never dismantle the master's house. They may allow us temporarily to beat him at his own game, but they will never enable us to bring about genuine change''~\cite{Lorde_1984}. Audre Lorde delivered these famous words to a conference of white feminist academics, pointing out how they were ostensibly advocating feminism but doing so with and within the patriarchal structures of academia. Her words are read to ``challenge reformist feminists to become radical feminists''~\cite{Olson_2000}, imploring them to lay down the allegorical tools of the master that may give them academic credibility while simultaneously holding down others who do not have access to these tools. Lorde asks: ``What does it mean when the tools of a racist patriarchy are used to examine the fruits of that same patriarchy?'' I ask an analogous question: if AI ethics relies on quantified neutral objectivity for its legitimacy, what will that foreclose? And Lorde responds: ``It means that only the most narrow perimeters of change are possible and allowable.'' The simple point of my piece is analogous: we reach a limit on the legitimacy that AI ethics can achieve while understanding itself as neutral and objective, which will thereby delegitimize practices of AI ethics that authentically integrate lived, located and embodied experiences. Put even more directly, by entrenching the epistemic power of quantification, the wide use of Model Cards and their ilk risk legitimizing \textit{AI ethics} as a project in equal and opposite measure to the extent that they delegitimize and marginalize embodied and lived experiences as legitimate parts of the same project. 

Using the metaphor of sight, Haraway asks us to reject the ``deadly fantasy that feminists and others have identified in \textit{some versions of objectivity}, those in the service of hierarchical and positivist orderings of what can count as knowledge''~\cite[p. 580, emphasis mine]{Haraway_1988}. Crucially, she does not give up on objectivity wholesale, and instead asks us to construct a feminist objectivity that is grounded in ``situated and embodied knowledges and an argument against various forms of unlocatable, and so irresponsible, knowledge claims''~\cite[p. 583]{Haraway_1988}.\footnote{Haraway's notion of situated knowledges is similar to the Harding discussion of strong objectivity~\cite{Harding_1992}. Haraway also emphasizes that she has no patience for relativism, calling it equally in denial of the ``stakes in location, embodiment, and partial perspective'' as both relativism and ``view from nowhere'' objectivity ``make it impossible to see well''~\cite[p. 584]{Haraway_1988}. Instead, she suggests how leaning in to the ``particularity and embodiment of all vision'' thus ``allows us to construct a usable, but not innocent ... objectivity''' (582). 
} In this way, feminist objectivity makes clear how objectivity is stronger when it acknowledges that all knowledge is socially situated, and the ways in which lived experience builds and inflects knowledge. And further, it is not quantification, \textit{per se}, that are the ``masters tools,'' instead it is the doctrinal positivist objectivity which disguises a hegemonic perspective as a ``view from nowhere,'' but which often adopts quantification as its tool.\footnote{Others have pointed out that an objectivity rooted in positivism is not commensurable with interpretivism given deep epistemological disagreements, even while their frequently chosen methods (ie, quantitative vs qualitative) may be commensurable~\cite[p. 174]{Lincoln_Guba_2000}.} 
Indeed, while examples of limited or oppressive uses of quantification abound~\cite{Browne_2015, Whittaker_2023, narayanan2022limits}, we can also find examples of interpretive quantification practices calculated toward better understanding one's embodied experience~\cite{Nafus_Sherman_2014}, documenting wider trends of oppression~\cite{D_Ignazio_2024, Suresh2022}, phenomena which affect our lives but occur at a scale too large to be appreciated directly~\cite{Ottinger_Bronson_Nafus_2023}, as well as other uses of quantification to disrupt power relations or advocate for the marginalized~\cite{Abebe_Barocas_Kleinberg_Levy_Raghavan_Robinson_2020, Gürses_Overdorf_Balsa_2018, Kulynych_Overdorf_Troncoso_Gürses_2020,keyes2022artificial}. This warning against entrenching the epistemic supremacy of quantification applies to these examples too: not to suggest that quantification must be eschewed, but instead to argue that steps must be taken to limit its epistemic supremacy in order to make space for other ways of knowing. 

Therefore, and perhaps unsurprisingly to feminist epistemologists, I label the ``view from nowhere,'' with its false guise of neutral objectivity, and the epistemic power that comes with it, as the master's tools. In her 2021 talk, Gebru talks about how epistemologies of AI adopt this ``view from nowhere,'' and use it to construct hierarchies of knowledge in AI. She relays how some receive who heard her proposal to include sociocultural understanding as part of ML data collection~\cite{Jo_Gebru_2020} responded to with the dismissive question ``What was the algorithmic component of her work?''---revealing the practice of improving an algorithmic technique on a quantifiable dataset to be the esteemed ``pinnacle''~\cite{Gebru_2021, gururaja2023build}. This question does much to disclose what is, and isn't, valued in the knowledge system of the dominant technical practices of today, and demonstrates how ``view from nowhere'' algorithms are given epistemic power. Relatedly, in an educational context, Raji, Scheuerman, and Amironesei demonstrate how courses teaching ``AI ethics'' often construe ethics using the language and techniques of computer science, reinforcing the discipline's dominance in hierarchies of knowledge, and thus resulting in an ``exclusionary pedagogy'' not open to other ways of knowing~\cite{raji2021you}. 

Black feminist philosopher Kristie Dotson reminds us how difficult of a road it will be to undo these hierarchies, on the road towards feminist objectivity that includes embodied and situated experiences. In her paper, \textit{Conceptualizing Epistemic Oppression}~\cite{Dotson_2014}, she expands beyond work on epistemic injustice~\cite{Fricker_2007} to remind us that the nature of epistemic power means that those holding it are very rarely confronted with the need to reevaluate their epistemic frames, and are therefore able to ``absorb extraordinarily large disturbances without redefining [their] structure.'' On the other hand, those using non-dominant epistemic systems may be ``rejected as nonsensical; [or as] as a deceiver with dangerous ideas; or [their] conclusions [..] might even invoke ridicule and laughter,'' or simply ``deep, awkward silences'' when scientists are confronted with indigenous knowledge claims they believe are objectively false~\cite[p. 143]{Verran_2013}. Dotson suggests that while ``Attempts to convince those who are relatively more epistemically powerful to relinquish some of their power might be warranted as a first line of address,'' one might also seek to lessen the effect of faulty credibility assessments. 

So, this difficult road implies the need for epistemological pluralism within quantitative or technical practice, in order to enable the incorporation and privileging of other kinds of knowledge, a project related to those others have attempted. Notably, we have Philip Agre's \textit{Toward a Critical Technical Practice: Lessons Learned in Trying to Reform AI} in which he relays his awakening out of dominant AI practices of ``write programs that solve problems better than anybody else's'' as the only valid proof of the veracity of a critique~\cite{Agre_1997}, and thus the only path to new knowledge, which nowadays usually means quantifying improvement on narrow benchmarks~\cite{gururaja2023build}. He critiques AI, but argues for the need to reform AI, into a practice that is critical and reflexive: a critical technical practice, asking us not to let go of the technical, but to be reflexively critical of its limits. More recently, Malik and Malik theorize ``Critical Technical Awakenings'' using Paulo Freire's theory of critical consciousness~\cite{Freire_1974} and emphasize the role of community, especially communities including non-technical and non-positivist thinkers, to enable more such awakenings~\cite{Malik_Malik_2021}. This aligns with calls to resist solitary and epistemically homogeneous approaches to teaching tech ethics~\cite{raji2021you}. Among other issues, though a process Ahmed calls ``atomization''~\cite[p. 180]{Ahmed_2021}, solitary approaches reduce the ability of a community to fashion and legitimize alternative epistemic practices: seen recently in how Microsoft disbanded a central ethics team and moved the few ethics practitioners who remained to report to technical product teams~\cite{Schiffer_Newton_2023}. Also, notably, there are common pitfalls on the road to epistemological pluralism, such as ``participatory'' AI approaches. Participatory AI falls short when those asked to ``participate'' are not given a say in the terms (including epistemic terms) of their participation nor a commitment to their ongoing inclusion~\cite{Sloane_Moss_Awomolo_Forlano_2022}, when careful participation is sacrificed in service of system scale~\cite{Delgado_Yang_Madaio_Yang_2023}, and when ``intersectional'' approaches to AI reduce the rich lived experience of intersectional oppression to subgroup parity fairness metrics~\cite{Kong_2022}.

Taking Agre's cogent cultural critique into an understanding of epistemic power, I ask those practicing ``AI,'' and other related quantified sciences widely practiced as value neutral, to help level epistemic power differentials by underscoring the partiality of their work~\cite{Suchman_2002}, constructing what I call \textit{humble technical practices}, which specifically seek to make their epistemic limits clear. Beyond the minimum technical humility sometimes demanded in ``threats to validity'' or ``limitations'' paper sections, which seek to enumerate and answer for possible technical flaws from within the epistemic frame of the paper, this would include gesturing towards (as doing more would likely be seen as nonsensical, until such practices become routine~\cite{Verran_2013}) the epistemic limits of quantification, and what other epistemic frames may have to offer. Beyond written products, this may be easier in interactive settings such as research talks or conversations with peers, and especially effective when given on prominent platforms. This would involve acknowledging the incomplete view that large-scale data analysis offers, and directing readers or the audience to other forms of knowledge on similar problems. This can also blend well with approaches to share epistemic power: elevating and platforming those with lived experience to serve as the natural expert on the nature of the problems they face and the roots of these problems. This should involve not only attention to citational justice by seeking to give epistemic power to marginalized academics and marginalized academic disciplines through citational practices~\cite{Kwon_2022}, but crucially must also seek to more broadly recognize knowledge sources not represented in tech companies or the academy. When lived experience is segmented off into its own conference track or side meetings~\cite{acmb}, it can literally be difficult to see them as worthy of being in the same room~\cite{gangadharan2020crafting}. 

Epistemic power is bound up with, and often flowing from, other kinds including economic and political power. For example, Forsythe writes how the (poor) appropriation of ethnographic methods in positivist ways by untrained AI researchers had economic outcomes: ``my own research funds have dried up''~\cite[p. 143]{Forsythe_2001}. It is therefore those with other kinds of power, along lines of race, gender, or wealth, that should first share or cede epistemic power. Notably, people harmed by AI are often not paid nor recognized for sharing their lived experience of this harm in the ways that academics may be paid to collect it~\cite{Sloane_Moss_Awomolo_Forlano_2022}. Therefore, I call for more ways to legitimize the sharing of lived experience in and on its own terms as valid AI ethics practice, and indeed, the preeminent and most true form of such practice. This must be listened to, and AI Ethics must be seen as a project to legitimize and amplify these on their own terms.

Technology audits led by advocates and activists, instead of those occurring as ``closed-door compliance'' exercises locked within technology companies and their epistemic frame, are a way to open the doors to other kinds of knowledge and thereby flatten epistemic power hierarchies. Peaks Krafft and coauthors propose a flowchart, questionnaire, and other tools to be wielded by ``those with the lived experience of being differentially targeted by surveillance technologies,'' specifically positioning this as \textit{expertise} of the sort particularly integral to equity in public sector algorithmic systems~\cite{Krafft_2021}. They conclude their work with the sendoff: ``The technofuture we project through this work is defiance rather than compliance,'' fitting given that their audit framework is designed to enable such defiance to be taken seriously. I echo this call for defiance: we will build the power we need when we speak from our experience, building ``collective knowledge of the specific locations of our specific visions''~\cite{Suchman_2002}, without hesitation and without reaching for other sources of legitimacy.

\section*{Acknowledgments}
This paper began as an answer to an incisive question that Dawn Nafus asked during my dissertation defense. 
Versions of this draft benefited from her generous comments and those of Blair Attard-Frost, Sireesh Gururaja, Benjamin Laufer, Lucy Suchman, Michael Madaio, and Os Keyes. 
Walks discussing these ideas with Maria Ryabova, Harini Suresh, Emily Tseng, and Frank Pasquale helped sharpen and encourage this work. 
I thank the Visual AI and HCI labs of Princeton CS for the opportunity to present early versions of this work, particularly Sunnie Kim, Andr\'es Monroy-Hern\'andez, and Olga Russakovsky. 
The early stages of writing this article were coincident with writing a forthcoming position paper written with Claire Le Goues on a similar topic in Software Engineering~\cite{widder2024whata}. 
Portions of data presented here were collected in collaboration with Derrick Zhen, Laura Dabbish, and Jim Herbsleb. 
My thinking has benefited from the intellectually diverse atmospheres of the Digital Life Initiative at Cornell Tech, the Feminist Political Economy reading group (open to all, join us!), and in my work with activist groups in Pittsburgh including Against Carceral Tech, the Coalition Against Predictive Policing, and \#NoTechForICE. 
May Carnegie Mellon never again deploy racist predictive policing~\cite{capp2018predictive} nor propose IoT algorithmic management of custodial workers~\cite{guo2023computer}, and learn from the lived experience of those such systems harm and the activists who fight against them.

\bibliographystyle{ACM-Reference-Format}
\bibliography{sample-base}


\begin{thebibliography}{102}


\ifx \showCODEN    \undefined \def \showCODEN     #1{\unskip}     \fi
\ifx \showDOI      \undefined \def \showDOI       #1{#1}\fi
\ifx \showISBNx    \undefined \def \showISBNx     #1{\unskip}     \fi
\ifx \showISBNxiii \undefined \def \showISBNxiii  #1{\unskip}     \fi
\ifx \showISSN     \undefined \def \showISSN      #1{\unskip}     \fi
\ifx \showLCCN     \undefined \def \showLCCN      #1{\unskip}     \fi
\ifx \shownote     \undefined \def \shownote      #1{#1}          \fi
\ifx \showarticletitle \undefined \def \showarticletitle #1{#1}   \fi
\ifx \showURL      \undefined \def \showURL       {\relax}        \fi
\providecommand\bibfield[2]{#2}
\providecommand\bibinfo[2]{#2}
\providecommand\natexlab[1]{#1}
\providecommand\showeprint[2][]{arXiv:#2}

\bibitem[acm(2023)]%
        {acmb}
 \bibinfo{year}{2023}\natexlab{}.
\newblock \bibinfo{title}{{{ACM FAccT}} 2023 {{Call}} for {{CRAFT Proposals}}}.
\newblock \bibinfo{howpublished}{https://facctconference.org/2023/cfpcraft}.
\newblock


\bibitem[acm(2024)]%
        {acm}
 \bibinfo{year}{2024}\natexlab{}.
\newblock \bibinfo{title}{{{ACM FAccT}} - 2024 {{Call for Papers}}}.
\newblock \bibinfo{howpublished}{https://facctconference.org/2024/cfp}.
\newblock


\bibitem[Abdurahman(2021)]%
        {abdurahman2021calculating}
\bibfield{author}{\bibinfo{person}{J.~Khadijah Abdurahman}.} \bibinfo{year}{2021}\natexlab{}.
\newblock \showarticletitle{Calculating the {{Souls}} of {{Black Folk}}: {{Predictive Analytics}} in the {{New York City Administration}} for {{Children}}'s {{Services}}}.
\newblock \bibinfo{journal}{\emph{Columbia Journal of Race and Law}}  \bibinfo{volume}{11} (\bibinfo{year}{2021}), \bibinfo{pages}{75--110}.
\newblock
\urldef\tempurl%
\url{https://journals.library.columbia.edu/index.php/cjrl/article/view/8741}
\showURL{%
\tempurl}


\bibitem[Abebe et~al\mbox{.}(2020)]%
        {Abebe_Barocas_Kleinberg_Levy_Raghavan_Robinson_2020}
\bibfield{author}{\bibinfo{person}{Rediet Abebe}, \bibinfo{person}{Solon Barocas}, \bibinfo{person}{Jon Kleinberg}, \bibinfo{person}{Karen Levy}, \bibinfo{person}{Manish Raghavan}, {and} \bibinfo{person}{David~G. Robinson}.} \bibinfo{year}{2020}\natexlab{}.
\newblock \showarticletitle{Roles for computing in social change}. In \bibinfo{booktitle}{\emph{Conference on Fairness, Accountability, and Transparency}} \emph{(\bibinfo{series}{FAT* ’20})}. \bibinfo{publisher}{Association for Computing Machinery}, \bibinfo{address}{New York, NY, USA}, \bibinfo{pages}{252–260}.
\newblock
\showISBNx{978-1-4503-6936-7}
\urldef\tempurl%
\url{https://doi.org/10.1145/3351095.3372871}
\showDOI{\tempurl}


\bibitem[Agre(1997)]%
        {Agre_1997}
\bibfield{author}{\bibinfo{person}{Philip Agre}.} \bibinfo{year}{1997}\natexlab{}.
\newblock \showarticletitle{Toward a Critical Technical Practice: Lessons Learned in Trying to Reform AI}.
\newblock In \bibinfo{booktitle}{\emph{Bridging the Great Divide: Social Science, Technical Systems, and Cooperative Work}}. \bibinfo{publisher}{Erlbaum}.
\newblock
\urldef\tempurl%
\url{https://pages.gseis.ucla.edu/faculty/agre/critical.html}
\showURL{%
\tempurl}


\bibitem[Ahmed(2021)]%
        {Ahmed_2021}
\bibfield{author}{\bibinfo{person}{Sara Ahmed}.} \bibinfo{year}{2021}\natexlab{}.
\newblock \bibinfo{booktitle}{\emph{Complaint!}}
\newblock \bibinfo{publisher}{Duke University Press}, \bibinfo{address}{Durham, NC}.
\newblock
\showISBNx{978-1-4780-1771-4}


\bibitem[Ajder et~al\mbox{.}(2019)]%
        {ajder2019state}
\bibfield{author}{\bibinfo{person}{Henry Ajder}, \bibinfo{person}{Giorgio Patrini}, \bibinfo{person}{Francesco Cavalli}, {and} \bibinfo{person}{Laurence Cullen}.} \bibinfo{year}{2019}\natexlab{}.
\newblock \showarticletitle{The State of Deepfakes: {{Landscape}}, Threats, and Impact}.
\newblock \bibinfo{journal}{\emph{Amsterdam: Deeptrace}} (\bibinfo{year}{2019}).
\newblock


\bibitem[Amugongo et~al\mbox{.}(2023)]%
        {amugongo2023invigorating}
\bibfield{author}{\bibinfo{person}{Lameck~Mbangula Amugongo}, \bibinfo{person}{Nicola~J. Bidwell}, {and} \bibinfo{person}{Caitlin~C. Corrigan}.} \bibinfo{year}{2023}\natexlab{}.
\newblock \showarticletitle{Invigorating {{Ubuntu Ethics}} in {{AI}} for Healthcare: {{Enabling}} Equitable Care}. In \bibinfo{booktitle}{\emph{2023 {{ACM Conference}} on {{Fairness}}, {{Accountability}}, and {{Transparency}}}}. \bibinfo{publisher}{{ACM}}, \bibinfo{address}{{Chicago IL USA}}, \bibinfo{pages}{583--592}.
\newblock
\showISBNx{9798400701924}
\urldef\tempurl%
\url{https://doi.org/10.1145/3593013.3594024}
\showDOI{\tempurl}


\bibitem[Andreessen(2023)]%
        {Andreessen_2023}
\bibfield{author}{\bibinfo{person}{Marc Andreessen}.} \bibinfo{year}{2023}\natexlab{}.
\newblock \bibinfo{title}{The Techno-Optimist Manifesto}.
\newblock
\newblock
\urldef\tempurl%
\url{https://a16z.com/the-techno-optimist-manifesto/}
\showURL{%
\tempurl}


\bibitem[Attard-Frost et~al\mbox{.}(2022)]%
        {Attard-Frost_De_los_Rios_Walters_2022}
\bibfield{author}{\bibinfo{person}{Blair Attard-Frost}, \bibinfo{person}{Andrés De~los Ríos}, {and} \bibinfo{person}{Deneille~R. Walters}.} \bibinfo{year}{2022}\natexlab{}.
\newblock \showarticletitle{The ethics of AI business practices: a review of 47 AI ethics guidelines}.
\newblock \bibinfo{journal}{\emph{AI and Ethics}} (\bibinfo{date}{April} \bibinfo{year}{2022}).
\newblock
\showISSN{2730-5953, 2730-5961}
\urldef\tempurl%
\url{https://doi.org/10.1007/s43681-022-00156-6}
\showDOI{\tempurl}


\bibitem[Bender et~al\mbox{.}(2021)]%
        {Bender_Gebru_McMillan-Major_Shmitchell_2021}
\bibfield{author}{\bibinfo{person}{Emily~M. Bender}, \bibinfo{person}{Timnit Gebru}, \bibinfo{person}{Angelina McMillan-Major}, {and} \bibinfo{person}{Shmargaret Shmitchell}.} \bibinfo{year}{2021}\natexlab{}.
\newblock \showarticletitle{On the Dangers of Stochastic Parrots: Can Language Models Be Too Big?}. In \bibinfo{booktitle}{\emph{Conference on Fairness, Accountability, and Transparency}}. \bibinfo{publisher}{ACM}, \bibinfo{address}{Virtual Event Canada}, \bibinfo{pages}{610–623}.
\newblock
\showISBNx{978-1-4503-8309-7}
\urldef\tempurl%
\url{https://doi.org/10.1145/3442188.3445922}
\showDOI{\tempurl}


\bibitem[Benjamin(2022)]%
        {benjamin2022fuckthealgorithm}
\bibfield{author}{\bibinfo{person}{Garfield Benjamin}.} \bibinfo{year}{2022}\natexlab{}.
\newblock \showarticletitle{\#{{FuckTheAlgorithm}}: Algorithmic Imaginaries and Political Resistance}. In \bibinfo{booktitle}{\emph{2022 {{ACM Conference}} on {{Fairness}}, {{Accountability}}, and {{Transparency}}}}. \bibinfo{publisher}{{ACM}}, \bibinfo{address}{{Seoul Republic of Korea}}, \bibinfo{pages}{46--57}.
\newblock
\showISBNx{978-1-4503-9352-2}
\urldef\tempurl%
\url{https://doi.org/10.1145/3531146.3533072}
\showDOI{\tempurl}


\bibitem[Birhane et~al\mbox{.}(2022)]%
        {birhane2022forgottena}
\bibfield{author}{\bibinfo{person}{Abeba Birhane}, \bibinfo{person}{Elayne Ruane}, \bibinfo{person}{Thomas Laurent}, \bibinfo{person}{Matthew S.~Brown}, \bibinfo{person}{Johnathan Flowers}, \bibinfo{person}{Anthony Ventresque}, {and} \bibinfo{person}{Christopher L.~Dancy}.} \bibinfo{year}{2022}\natexlab{}.
\newblock \showarticletitle{The {{Forgotten Margins}} of {{AI Ethics}}}. In \bibinfo{booktitle}{\emph{2022 {{ACM Conference}} on {{Fairness}}, {{Accountability}}, and {{Transparency}}}} \emph{(\bibinfo{series}{{{FAccT}} '22})}. \bibinfo{publisher}{{Association for Computing Machinery}}, \bibinfo{address}{{New York, NY, USA}}, \bibinfo{pages}{948--958}.
\newblock
\showISBNx{978-1-4503-9352-2}
\urldef\tempurl%
\url{https://doi.org/10.1145/3531146.3533157}
\showDOI{\tempurl}


\bibitem[Boag et~al\mbox{.}(2022)]%
        {boag2022techa}
\bibfield{author}{\bibinfo{person}{William Boag}, \bibinfo{person}{Harini Suresh}, \bibinfo{person}{Bianca Lepe}, {and} \bibinfo{person}{Catherine D'Ignazio}.} \bibinfo{year}{2022}\natexlab{}.
\newblock \showarticletitle{Tech {{Worker Organizing}} for {{Power}} and {{Accountability}}}. In \bibinfo{booktitle}{\emph{2022 {{ACM Conference}} on {{Fairness}}, {{Accountability}}, and {{Transparency}}}}. \bibinfo{publisher}{{ACM}}, \bibinfo{address}{{Seoul Republic of Korea}}, \bibinfo{pages}{452--463}.
\newblock
\showISBNx{978-1-4503-9352-2}
\urldef\tempurl%
\url{https://doi.org/10.1145/3531146.3533111}
\showDOI{\tempurl}


\bibitem[Braun and Clarke(2006)]%
        {Braun_Clarke_2006}
\bibfield{author}{\bibinfo{person}{Virginia Braun} {and} \bibinfo{person}{Victoria Clarke}.} \bibinfo{year}{2006}\natexlab{}.
\newblock \showarticletitle{Using thematic analysis in psychology}.
\newblock \bibinfo{journal}{\emph{Qualitative Research in Psychology}} \bibinfo{volume}{3}, \bibinfo{number}{2} (\bibinfo{date}{Jan.} \bibinfo{year}{2006}), \bibinfo{pages}{77–101}.
\newblock
\showISSN{1478-0887}
\urldef\tempurl%
\url{https://doi.org/10.1191/1478088706qp063oa}
\showDOI{\tempurl}


\bibitem[Braun and Clarke(2019)]%
        {Braun_Clarke_2019}
\bibfield{author}{\bibinfo{person}{Virginia Braun} {and} \bibinfo{person}{Victoria Clarke}.} \bibinfo{year}{2019}\natexlab{}.
\newblock \showarticletitle{Reflecting on reflexive thematic analysis}.
\newblock \bibinfo{journal}{\emph{Qualitative Research in Sport, Exercise and Health}} \bibinfo{volume}{11}, \bibinfo{number}{4} (\bibinfo{date}{Aug.} \bibinfo{year}{2019}), \bibinfo{pages}{589–597}.
\newblock
\showISSN{2159-676X, 2159-6778}
\urldef\tempurl%
\url{https://doi.org/10.1080/2159676X.2019.1628806}
\showDOI{\tempurl}


\bibitem[Braun and Clarke(2021)]%
        {braun2021one}
\bibfield{author}{\bibinfo{person}{Virginia Braun} {and} \bibinfo{person}{Victoria Clarke}.} \bibinfo{year}{2021}\natexlab{}.
\newblock \showarticletitle{One Size Fits All? {{What}} Counts as Quality Practice in (Reflexive) Thematic Analysis?}
\newblock \bibinfo{journal}{\emph{Qualitative Research in Psychology}}  \bibinfo{volume}{18} (\bibinfo{year}{2021}), \bibinfo{pages}{328--352}.
\newblock
\urldef\tempurl%
\url{https://www.tandfonline.com/doi/full/10.1080/14780887.2020.1769238}
\showURL{%
\tempurl}


\bibitem[Broussard(2019)]%
        {Broussard_2019}
\bibfield{author}{\bibinfo{person}{Meredith Broussard}.} \bibinfo{year}{2019}\natexlab{}.
\newblock \bibinfo{booktitle}{\emph{Statement before the Task Force on Artificial Intelligence of the Committee on Financial Services, U.S. House of Representatives}}.
\newblock


\bibitem[Browne(2015)]%
        {Browne_2015}
\bibfield{author}{\bibinfo{person}{Simone Browne}.} \bibinfo{year}{2015}\natexlab{}.
\newblock \bibinfo{booktitle}{\emph{Dark Matters: On the Surveillance of Blackness}}.
\newblock \bibinfo{publisher}{Duke University Press}.
\newblock
\showISBNx{978-0-8223-7530-2}
\urldef\tempurl%
\url{https://doi.org/10.1215/9780822375302}
\showDOI{\tempurl}


\bibitem[Buolamwini and Gebru(2018)]%
        {Buolamwini_Gebru_2018}
\bibfield{author}{\bibinfo{person}{Joy Buolamwini} {and} \bibinfo{person}{Timnit Gebru}.} \bibinfo{year}{2018}\natexlab{}.
\newblock \showarticletitle{Gender shades: Intersectional accuracy disparities in commercial gender classification}. In \bibinfo{booktitle}{\emph{Conference on fairness, accountability, and transparency}}. \bibinfo{publisher}{Proceedings of Machine Learning Research}, \bibinfo{pages}{77–91}.
\newblock


\bibitem[Coalition Against Predictive~Policing(2018)]%
        {capp2018predictive}
\bibfield{author}{\bibinfo{person}{Pittsburgh Coalition Against Predictive~Policing}.} \bibinfo{year}{2018}\natexlab{}.
\newblock \showarticletitle{Predictive Policing in Pittsburgh: A Primer}.
\newblock  (\bibinfo{year}{2018}).
\newblock
\urldef\tempurl%
\url{https://capp-pgh.com/files/Primer_v1.pdf}
\showURL{%
\tempurl}


\bibitem[Cobbe et~al\mbox{.}(2023)]%
        {cobbe2023understandinga}
\bibfield{author}{\bibinfo{person}{Jennifer Cobbe}, \bibinfo{person}{Michael Veale}, {and} \bibinfo{person}{Jatinder Singh}.} \bibinfo{year}{2023}\natexlab{}.
\newblock \showarticletitle{Understanding Accountability in Algorithmic Supply Chains}. In \bibinfo{booktitle}{\emph{2023 {{ACM Conference}} on {{Fairness}}, {{Accountability}}, and {{Transparency}}}}. \bibinfo{publisher}{{ACM}}, \bibinfo{address}{{Chicago IL USA}}, \bibinfo{pages}{1186--1197}.
\newblock
\showISBNx{9798400701924}
\urldef\tempurl%
\url{https://doi.org/10.1145/3593013.3594073}
\showDOI{\tempurl}


\bibitem[Daston and Galison(2007)]%
        {Daston_Galison_2007}
\bibfield{author}{\bibinfo{person}{Lorraine Daston} {and} \bibinfo{person}{Peter~Louis Galison}.} \bibinfo{year}{2007}\natexlab{}.
\newblock \bibinfo{booktitle}{\emph{Objectivity}}.
\newblock \bibinfo{publisher}{Zone Books}, \bibinfo{address}{New York}.
\newblock
\showISBNx{978-1-890951-78-8}


\bibitem[Davidson(2020)]%
        {davidson2020pittsburgh}
\bibfield{author}{\bibinfo{person}{Tom Davidson}.} \bibinfo{year}{2020}\natexlab{}.
\newblock \showarticletitle{Pittsburgh {{City Council}} Told Facial Recognition Bill Should Be Stronger}.
\newblock \bibinfo{journal}{\emph{TribLIVE}} (\bibinfo{date}{Sept.} \bibinfo{year}{2020}).
\newblock


\bibitem[Delgado et~al\mbox{.}(2023)]%
        {Delgado_Yang_Madaio_Yang_2023}
\bibfield{author}{\bibinfo{person}{Fernando Delgado}, \bibinfo{person}{Stephen Yang}, \bibinfo{person}{Michael Madaio}, {and} \bibinfo{person}{Qian Yang}.} \bibinfo{year}{2023}\natexlab{}.
\newblock \showarticletitle{The Participatory Turn in AI Design: Theoretical Foundations and the Current State of Practice}. In \bibinfo{booktitle}{\emph{Conference on Equity and Access in Algorithms, Mechanisms, and Optimization}} \emph{(\bibinfo{series}{EAAMO ’23})}. \bibinfo{publisher}{Association for Computing Machinery}, \bibinfo{address}{New York, NY, USA}, \bibinfo{pages}{1–23}.
\newblock
\showISBNx{9798400703812}
\urldef\tempurl%
\url{https://doi.org/10.1145/3617694.3623261}
\showDOI{\tempurl}


\bibitem[Deng et~al\mbox{.}(2023)]%
        {deng2023investigating}
\bibfield{author}{\bibinfo{person}{Wesley~Hanwen Deng}, \bibinfo{person}{Nur Yildirim}, \bibinfo{person}{Monica Chang}, \bibinfo{person}{Motahhare Eslami}, \bibinfo{person}{Kenneth Holstein}, {and} \bibinfo{person}{Michael Madaio}.} \bibinfo{year}{2023}\natexlab{}.
\newblock \showarticletitle{Investigating {{Practices}} and {{Opportunities}} for {{Cross-functional Collaboration}} around {{AI Fairness}} in {{Industry Practice}}}. In \bibinfo{booktitle}{\emph{2023 {{ACM Conference}} on {{Fairness}}, {{Accountability}}, and {{Transparency}}}}. \bibinfo{publisher}{{ACM}}, \bibinfo{address}{{Chicago IL USA}}, \bibinfo{pages}{705--716}.
\newblock
\showISBNx{9798400701924}
\urldef\tempurl%
\url{https://doi.org/10.1145/3593013.3594037}
\showDOI{\tempurl}


\bibitem[Dictionary(2023a)]%
        {Oxford_English_Dictionary_2023b}
\bibfield{author}{\bibinfo{person}{Oxford~English Dictionary}.} \bibinfo{year}{2023}\natexlab{a}.
\newblock \bibinfo{title}{objective, adj., sense I.3.b}.
\newblock
\newblock
\urldef\tempurl%
\url{https://doi.org/10.1093/OED/3908623815}
\showURL{%
\tempurl}


\bibitem[Dictionary(2023b)]%
        {Oxford_English_Dictionary_2023a}
\bibfield{author}{\bibinfo{person}{Oxford~English Dictionary}.} \bibinfo{year}{2023}\natexlab{b}.
\newblock \bibinfo{title}{subjective, adj., sense 3}.
\newblock
\newblock
\urldef\tempurl%
\url{https://doi.org/10.1093/OED/1098712908}
\showURL{%
\tempurl}


\bibitem[Dotson(2014)]%
        {Dotson_2014}
\bibfield{author}{\bibinfo{person}{Kristie Dotson}.} \bibinfo{year}{2014}\natexlab{}.
\newblock \showarticletitle{Conceptualizing Epistemic Oppression}.
\newblock \bibinfo{journal}{\emph{Social Epistemology}} \bibinfo{volume}{28}, \bibinfo{number}{2} (\bibinfo{date}{April} \bibinfo{year}{2014}), \bibinfo{pages}{115–138}.
\newblock
\showISSN{0269-1728}
\urldef\tempurl%
\url{https://doi.org/10.1080/02691728.2013.782585}
\showDOI{\tempurl}


\bibitem[D’Ignazio(2024)]%
        {D_Ignazio_2024}
\bibfield{author}{\bibinfo{person}{Catherine D’Ignazio}.} \bibinfo{year}{2024}\natexlab{}.
\newblock \bibinfo{booktitle}{\emph{Counting Feminicide: Data Feminism in Action}}.
\newblock \bibinfo{publisher}{MIT Press}.
\newblock
\showISBNx{978-0-262-37800-0}
\newblock
\shownote{Google-Books-ID: 2STOEAAAQBAJ}.


\bibitem[Edenberg and Wood(2023)]%
        {edenberg2023epistemic}
\bibfield{author}{\bibinfo{person}{Elizabeth Edenberg} {and} \bibinfo{person}{Alexandra Wood}.} \bibinfo{year}{2023}\natexlab{}.
\newblock \showarticletitle{An {{Epistemic Lens}} on {{Algorithmic Fairness}}}. In \bibinfo{booktitle}{\emph{Equity and {{Access}} in {{Algorithms}}, {{Mechanisms}}, and {{Optimization}}}}. \bibinfo{publisher}{{ACM}}, \bibinfo{address}{{Boston MA USA}}, \bibinfo{pages}{1--10}.
\newblock
\showISBNx{9798400703812}
\urldef\tempurl%
\url{https://doi.org/10.1145/3617694.3623248}
\showDOI{\tempurl}


\bibitem[Forsythe(2001)]%
        {Forsythe_2001}
\bibfield{author}{\bibinfo{person}{Diana~E. Forsythe}.} \bibinfo{year}{2001}\natexlab{}.
\newblock \bibinfo{booktitle}{\emph{Studying Those Who Study Us: An Anthropologist in the World of Artificial Intelligence}}.
\newblock \bibinfo{publisher}{Stanford University Press}, \bibinfo{address}{Stanford}.
\newblock
\showISBNx{978-0-8047-4203-0}


\bibitem[Freire(1974)]%
        {Freire_1974}
\bibfield{author}{\bibinfo{person}{Paulo Freire}.} \bibinfo{year}{1974}\natexlab{}.
\newblock \bibinfo{booktitle}{\emph{Education for critical consciousness}}.
\newblock \bibinfo{publisher}{Continuum}, \bibinfo{address}{London}.
\newblock
\showISBNx{978-0-8264-7795-8}


\bibitem[Fricker(2007)]%
        {Fricker_2007}
\bibfield{author}{\bibinfo{person}{Miranda Fricker}.} \bibinfo{year}{2007}\natexlab{}.
\newblock \bibinfo{booktitle}{\emph{Epistemic injustice: power and the ethics of knowing}}.
\newblock \bibinfo{publisher}{Oxford University Press}, \bibinfo{address}{Oxford; New York}.
\newblock
\showISBNx{978-0-19-823790-7}


\bibitem[Gangadharan and Gürses(2020)]%
        {gangadharan2020crafting}
\bibfield{author}{\bibinfo{person}{Seeta~Peña Gangadharan} {and} \bibinfo{person}{Seda Gürses}.} \bibinfo{year}{2020}\natexlab{}.
\newblock \bibinfo{booktitle}{\emph{Crafting {{CRAFT}}}}.
\newblock
\urldef\tempurl%
\url{https://craft-workshop.gitlab.io/}
\showURL{%
\tempurl}


\bibitem[Gansky and McDonald(2022)]%
        {Gansky_McDonald_2022}
\bibfield{author}{\bibinfo{person}{Ben Gansky} {and} \bibinfo{person}{Sean McDonald}.} \bibinfo{year}{2022}\natexlab{}.
\newblock \showarticletitle{CounterFAccTual: How FAccT Undermines Its Organizing Principles}. In \bibinfo{booktitle}{\emph{Conference on Fairness, Accountability, and Transparency}}. \bibinfo{publisher}{ACM}, \bibinfo{address}{Seoul Republic of Korea}, \bibinfo{pages}{1982–1992}.
\newblock
\showISBNx{978-1-4503-9352-2}
\urldef\tempurl%
\url{https://doi.org/10.1145/3531146.3533241}
\showDOI{\tempurl}


\bibitem[Gebru(2021)]%
        {Gebru_2021}
\bibfield{author}{\bibinfo{person}{Timnit Gebru}.} \bibinfo{year}{2021}\natexlab{}.
\newblock \bibinfo{title}{Hierarchy of Knowledge in Machine Learning \& Related Fields \& Its Consequences}.
\newblock
\newblock
\urldef\tempurl%
\url{https://scs.hosted.panopto.com/Panopto/Pages/Viewer.aspx?id=70f6edd7-de91-464e-ae94-acbb011ba2c7}
\showURL{%
\tempurl}


\bibitem[Gebru et~al\mbox{.}(2021)]%
        {gebru2021datasheets}
\bibfield{author}{\bibinfo{person}{Timnit Gebru}, \bibinfo{person}{Jamie Morgenstern}, \bibinfo{person}{Briana Vecchione}, \bibinfo{person}{Jennifer~Wortman Vaughan}, \bibinfo{person}{Hanna Wallach}, \bibinfo{person}{Hal~Daum{\'e} Iii}, {and} \bibinfo{person}{Kate Crawford}.} \bibinfo{year}{2021}\natexlab{}.
\newblock \showarticletitle{Datasheets for Datasets}.
\newblock \bibinfo{journal}{\emph{Commun. ACM}}  \bibinfo{volume}{64} (\bibinfo{date}{Dec.} \bibinfo{year}{2021}), \bibinfo{pages}{86--92}.
\newblock


\bibitem[Google({[n.\,d.]})]%
        {Google}
\bibfield{author}{\bibinfo{person}{Google}.} \bibinfo{year}{[n.\,d.]}\natexlab{}.
\newblock \bibinfo{title}{Google Cloud Model Cards}.
\newblock
\newblock
\urldef\tempurl%
\url{https://modelcards.withgoogle.com/about}
\showURL{%
\tempurl}


\bibitem[Greene et~al\mbox{.}(2019)]%
        {Greene_Hoffmann_Stark_2019}
\bibfield{author}{\bibinfo{person}{Daniel Greene}, \bibinfo{person}{Anna~Lauren Hoffmann}, {and} \bibinfo{person}{Luke Stark}.} \bibinfo{year}{2019}\natexlab{}.
\newblock \showarticletitle{Better, Nicer, Clearer, Fairer: A Critical Assessment of the Movement for Ethical Artificial Intelligence and Machine Learning}.
\newblock \bibinfo{journal}{\emph{Hawaii International Conference on System Sciences 2019 (HICSS-52)}} (\bibinfo{date}{Jan.} \bibinfo{year}{2019}).
\newblock
\urldef\tempurl%
\url{https://aisel.aisnet.org/hicss-52/dsm/critical_and_ethical_studies/2}
\showURL{%
\tempurl}


\bibitem[Gulland and Tan(2023)]%
        {gulland2023campus}
\bibfield{author}{\bibinfo{person}{Anne Gulland} {and} \bibinfo{person}{Fayth Tan}.} \bibinfo{year}{2023}\natexlab{}.
\newblock \showarticletitle{Campus Surveillance: Students and Professors Decry Sensors in Buildings}.
\newblock \bibinfo{journal}{\emph{Nature}} \bibinfo{volume}{623}, \bibinfo{number}{7986} (\bibinfo{date}{Oct.} \bibinfo{year}{2023}), \bibinfo{pages}{444--445}.
\newblock
\urldef\tempurl%
\url{https://doi.org/10.1038/d41586-023-03287-w}
\showDOI{\tempurl}


\bibitem[Guo and {Ryan-Mosley}(2023)]%
        {guo2023computer}
\bibfield{author}{\bibinfo{person}{Eileen Guo} {and} \bibinfo{person}{Tate {Ryan-Mosley}}.} \bibinfo{year}{2023}\natexlab{}.
\newblock \bibinfo{title}{Computer Scientists Designing the Future Can't Agree on What Privacy Means}.
\newblock \bibinfo{howpublished}{https://www.technologyreview.com/2023/04/03/1070665/cmu-university-privacy-battle-smart-building-sensors-mites/}.
\newblock


\bibitem[Gururaja et~al\mbox{.}(2023)]%
        {gururaja2023build}
\bibfield{author}{\bibinfo{person}{Sireesh Gururaja}, \bibinfo{person}{Amanda Bertsch}, \bibinfo{person}{Clara Na}, \bibinfo{person}{David Widder}, {and} \bibinfo{person}{Emma Strubell}.} \bibinfo{year}{2023}\natexlab{}.
\newblock \showarticletitle{To {{Build Our Future}}, {{We Must Know Our Past}}: {{Contextualizing Paradigm Shifts}} in {{Natural Language Processing}}}. In \bibinfo{booktitle}{\emph{{{Conference}} on {{Empirical Methods}} in {{Natural Language Processing}}}}. \bibinfo{publisher}{Association for Computational Linguistics}, \bibinfo{address}{Singapore}, \bibinfo{pages}{13310--13325}.
\newblock


\bibitem[Gürses et~al\mbox{.}(2018)]%
        {Gürses_Overdorf_Balsa_2018}
\bibfield{author}{\bibinfo{person}{Seda Gürses}, \bibinfo{person}{Rebekah Overdorf}, {and} \bibinfo{person}{Ero Balsa}.} \bibinfo{year}{2018}\natexlab{}.
\newblock \showarticletitle{POTs: the revolution will not be optimized?}
\newblock \bibinfo{journal}{\emph{11th Hot Topics in Privacy Enhancing Technologies (HotPETs)}} (\bibinfo{year}{2018}), \bibinfo{pages}{2}.
\newblock


\bibitem[Hancox-Li and Kumar(2021)]%
        {hancox-li2021epistemica}
\bibfield{author}{\bibinfo{person}{Leif Hancox-Li} {and} \bibinfo{person}{I.~Elizabeth Kumar}.} \bibinfo{year}{2021}\natexlab{}.
\newblock \showarticletitle{Epistemic Values in Feature Importance Methods: {{Lessons}} from Feminist Epistemology}. In \bibinfo{booktitle}{\emph{{{Conference}} on {{Fairness}}, {{Accountability}}, and {{Transparency}}}}. \bibinfo{publisher}{ACM}, \bibinfo{pages}{817--826}.
\newblock
\urldef\tempurl%
\url{https://dl.acm.org/doi/10.1145/3442188.3445943}
\showURL{%
\tempurl}


\bibitem[Haraway(1988)]%
        {Haraway_1988}
\bibfield{author}{\bibinfo{person}{Donna Haraway}.} \bibinfo{year}{1988}\natexlab{}.
\newblock \showarticletitle{Situated Knowledges: The Science Question in Feminism and the Privilege of Partial Perspective}.
\newblock \bibinfo{journal}{\emph{Feminist Studies}} \bibinfo{volume}{14}, \bibinfo{number}{3} (\bibinfo{year}{1988}), \bibinfo{pages}{575}.
\newblock
\showISSN{00463663}
\urldef\tempurl%
\url{https://doi.org/10.2307/3178066}
\showDOI{\tempurl}


\bibitem[Harding(1992)]%
        {Harding_1992}
\bibfield{author}{\bibinfo{person}{Sandra Harding}.} \bibinfo{year}{1992}\natexlab{}.
\newblock \showarticletitle{Rethinking Standpoint Epistemology: What Is “Strong Objectivity?”}.
\newblock \bibinfo{journal}{\emph{The Centennial Review}} \bibinfo{volume}{36}, \bibinfo{number}{3} (\bibinfo{year}{1992}), \bibinfo{pages}{437–470}.
\newblock
\showISSN{0162-0177}


\bibitem[Jo and Gebru(2020)]%
        {Jo_Gebru_2020}
\bibfield{author}{\bibinfo{person}{Eun~Seo Jo} {and} \bibinfo{person}{Timnit Gebru}.} \bibinfo{year}{2020}\natexlab{}.
\newblock \showarticletitle{Lessons from archives: strategies for collecting sociocultural data in machine learning}. In \bibinfo{booktitle}{\emph{Conference on Fairness, Accountability, and Transparency}}. \bibinfo{publisher}{ACM}, \bibinfo{address}{Barcelona Spain}, \bibinfo{pages}{306–316}.
\newblock
\showISBNx{978-1-4503-6936-7}
\urldef\tempurl%
\url{https://doi.org/10.1145/3351095.3372829}
\showDOI{\tempurl}


\bibitem[Jobin et~al\mbox{.}(2019)]%
        {Jobin_Ienca_Vayena_2019}
\bibfield{author}{\bibinfo{person}{Anna Jobin}, \bibinfo{person}{Marcello Ienca}, {and} \bibinfo{person}{Effy Vayena}.} \bibinfo{year}{2019}\natexlab{}.
\newblock \showarticletitle{The global landscape of AI ethics guidelines}.
\newblock \bibinfo{journal}{\emph{Nature Machine Intelligence}} \bibinfo{volume}{1}, \bibinfo{number}{9} (\bibinfo{year}{2019}), \bibinfo{pages}{389–399}.
\newblock


\bibitem[Kasy and Abebe(2021)]%
        {kasy2021fairness}
\bibfield{author}{\bibinfo{person}{Maximilian Kasy} {and} \bibinfo{person}{Rediet Abebe}.} \bibinfo{year}{2021}\natexlab{}.
\newblock \showarticletitle{Fairness, {{Equality}}, and {{Power}} in {{Algorithmic Decision-Making}}}. In \bibinfo{booktitle}{\emph{{{Conference}} on {{Fairness}}, {{Accountability}}, and {{Transparency}}}}. \bibinfo{publisher}{{ACM}}, \bibinfo{address}{{Virtual Event Canada}}, \bibinfo{pages}{576--586}.
\newblock
\showISBNx{978-1-4503-8309-7}
\urldef\tempurl%
\url{https://doi.org/10.1145/3442188.3445919}
\showDOI{\tempurl}


\bibitem[Keyes and Creel(2022)]%
        {keyes2022artificial}
\bibfield{author}{\bibinfo{person}{Os Keyes} {and} \bibinfo{person}{Kathleen Creel}.} \bibinfo{year}{2022}\natexlab{}.
\newblock \showarticletitle{Artificial {{Knowing Otherwise}}}.
\newblock \bibinfo{journal}{\emph{Feminist Philosophy Quarterly}}  \bibinfo{volume}{8} (\bibinfo{year}{2022}).
\newblock


\bibitem[Kong(2022)]%
        {Kong_2022}
\bibfield{author}{\bibinfo{person}{Youjin Kong}.} \bibinfo{year}{2022}\natexlab{}.
\newblock \showarticletitle{Are “Intersectionally Fair” AI Algorithms Really Fair to Women of Color? A Philosophical Analysis}. In \bibinfo{booktitle}{\emph{Conference on Fairness, Accountability, and Transparency}} \emph{(\bibinfo{series}{FAccT ’22})}. \bibinfo{publisher}{Association for Computing Machinery}, \bibinfo{address}{New York, NY, USA}, \bibinfo{pages}{485–494}.
\newblock
\showISBNx{978-1-4503-9352-2}
\urldef\tempurl%
\url{https://doi.org/10.1145/3531146.3533114}
\showDOI{\tempurl}


\bibitem[Krafft et~al\mbox{.}(2021)]%
        {Krafft_2021}
\bibfield{author}{\bibinfo{person}{P.~M. Krafft}, \bibinfo{person}{Meg Young}, \bibinfo{person}{Michael Katell}, \bibinfo{person}{Jennifer~E. Lee}, \bibinfo{person}{Shankar Narayan}, \bibinfo{person}{Micah Epstein}, \bibinfo{person}{Dharma Dailey}, \bibinfo{person}{Bernease Herman}, \bibinfo{person}{Aaron Tam}, \bibinfo{person}{Vivian Guetler}, \bibinfo{person}{Corinne Bintz}, \bibinfo{person}{Daniella Raz}, \bibinfo{person}{Pa~Ousman Jobe}, \bibinfo{person}{Franziska Putz}, \bibinfo{person}{Brian Robick}, {and} \bibinfo{person}{Bissan Barghouti}.} \bibinfo{year}{2021}\natexlab{}.
\newblock \showarticletitle{An Action-Oriented AI Policy Toolkit for Technology Audits by Community Advocates and Activists}. In \bibinfo{booktitle}{\emph{Conference on Fairness, Accountability, and Transparency}} \emph{(\bibinfo{series}{FAccT ’21})}. \bibinfo{publisher}{Association for Computing Machinery}, \bibinfo{address}{New York, NY, USA}, \bibinfo{pages}{772–781}.
\newblock
\showISBNx{978-1-4503-8309-7}
\urldef\tempurl%
\url{https://doi.org/10.1145/3442188.3445938}
\showDOI{\tempurl}


\bibitem[Kubit(2022)]%
        {kubit2022cmu}
\bibfield{author}{\bibinfo{person}{Lilly Kubit}.} \bibinfo{year}{2022}\natexlab{}.
\newblock \showarticletitle{On {{CMU}}'s Campus, Protesters March against Facial Recognition Tech}.
\newblock \bibinfo{journal}{\emph{PublicSource}} (\bibinfo{date}{Sept.} \bibinfo{year}{2022}).
\newblock


\bibitem[Kulynych et~al\mbox{.}(2020)]%
        {Kulynych_Overdorf_Troncoso_Gürses_2020}
\bibfield{author}{\bibinfo{person}{Bogdan Kulynych}, \bibinfo{person}{Rebekah Overdorf}, \bibinfo{person}{Carmela Troncoso}, {and} \bibinfo{person}{Seda Gürses}.} \bibinfo{year}{2020}\natexlab{}.
\newblock \showarticletitle{POTs: protective optimization technologies}. In \bibinfo{booktitle}{\emph{Conference on Fairness, Accountability, and Transparency}}. \bibinfo{publisher}{ACM}, \bibinfo{address}{Barcelona Spain}, \bibinfo{pages}{177–188}.
\newblock
\showISBNx{978-1-4503-6936-7}
\urldef\tempurl%
\url{https://doi.org/10.1145/3351095.3372853}
\showDOI{\tempurl}


\bibitem[Kwon(2022)]%
        {Kwon_2022}
\bibfield{author}{\bibinfo{person}{Diana Kwon}.} \bibinfo{year}{2022}\natexlab{}.
\newblock \showarticletitle{The rise of citational justice: how scholars are making references fairer}.
\newblock \bibinfo{journal}{\emph{Nature}} \bibinfo{volume}{603}, \bibinfo{number}{7902} (\bibinfo{date}{March} \bibinfo{year}{2022}), \bibinfo{pages}{568–571}.
\newblock
\urldef\tempurl%
\url{https://doi.org/10.1038/d41586-022-00793-1}
\showDOI{\tempurl}


\bibitem[Latour(1993)]%
        {Latour_1993}
\bibfield{author}{\bibinfo{person}{Bruno Latour}.} \bibinfo{year}{1993}\natexlab{}.
\newblock \bibinfo{booktitle}{\emph{We have never been modern}}.
\newblock \bibinfo{publisher}{Harvard University Press}, \bibinfo{address}{Cambridge, Mass}.
\newblock
\showISBNx{978-0-674-94838-9}


\bibitem[Laufer et~al\mbox{.}(2022)]%
        {laufer2022four}
\bibfield{author}{\bibinfo{person}{Benjamin Laufer}, \bibinfo{person}{Sameer Jain}, \bibinfo{person}{A.~Feder Cooper}, \bibinfo{person}{Jon Kleinberg}, {and} \bibinfo{person}{Hoda Heidari}.} \bibinfo{year}{2022}\natexlab{}.
\newblock \showarticletitle{Four {{Years}} of {{FAccT}}: {{A Reflexive}}, {{Mixed-Methods Analysis}} of {{Research Contributions}}, {{Shortcomings}}, and {{Future Prospects}}}. In \bibinfo{booktitle}{\emph{2022 {{ACM Conference}} on {{Fairness}}, {{Accountability}}, and {{Transparency}}}}. \bibinfo{publisher}{{ACM}}, \bibinfo{address}{{Seoul Republic of Korea}}, \bibinfo{pages}{401--426}.
\newblock
\showISBNx{978-1-4503-9352-2}
\urldef\tempurl%
\url{https://doi.org/10.1145/3531146.3533107}
\showDOI{\tempurl}


\bibitem[Lincoln and Guba(2000)]%
        {Lincoln_Guba_2000}
\bibfield{author}{\bibinfo{person}{Yvonna~S Lincoln} {and} \bibinfo{person}{Egon~G Guba}.} \bibinfo{year}{2000}\natexlab{}.
\newblock \showarticletitle{Paradigmatic controversies, contradictions, and emerging confluences}.
\newblock \bibinfo{journal}{\emph{The Sage handbook of qualitative research}}  \bibinfo{volume}{4} (\bibinfo{year}{2000}), \bibinfo{pages}{163–188}.
\newblock


\bibitem[Lohr(2018)]%
        {Lohr_2018}
\bibfield{author}{\bibinfo{person}{Steve Lohr}.} \bibinfo{year}{2018}\natexlab{}.
\newblock \showarticletitle{Facial Recognition Is Accurate, if You’re a White Guy}.
\newblock \bibinfo{journal}{\emph{The New York Times}} (\bibinfo{date}{Feb.} \bibinfo{year}{2018}).
\newblock
\showISSN{0362-4331}
\urldef\tempurl%
\url{https://www.nytimes.com/2018/02/09/technology/facial-recognition-race-artificial-intelligence.html}
\showURL{%
\tempurl}


\bibitem[Lorde(1984)]%
        {Lorde_1984}
\bibfield{author}{\bibinfo{person}{Audre Lorde}.} \bibinfo{year}{1984}\natexlab{}.
\newblock \showarticletitle{The Master’s Tools Will Never Dismantle the Master’s House}.
\newblock In \bibinfo{booktitle}{\emph{Sister Outsider: Essays and Speeches}}. \bibinfo{publisher}{Crossing Press}, \bibinfo{address}{Berkeley, CA}, \bibinfo{pages}{110–114}.
\newblock
\urldef\tempurl%
\url{https://collectiveliberation.org/wp-content/uploads/2013/01/Lorde_The_Masters_Tools.pdf}
\showURL{%
\tempurl}


\bibitem[Madaio et~al\mbox{.}(2020)]%
        {Madaio_Stark_Wortman_Vaughan_Wallach_2020}
\bibfield{author}{\bibinfo{person}{Michael~A Madaio}, \bibinfo{person}{Luke Stark}, \bibinfo{person}{Jennifer Wortman~Vaughan}, {and} \bibinfo{person}{Hanna Wallach}.} \bibinfo{year}{2020}\natexlab{}.
\newblock \showarticletitle{Co-designing checklists to understand organizational challenges and opportunities around fairness in ai}. In \bibinfo{booktitle}{\emph{Conference on Human Factors in Computing Systems}}. \bibinfo{publisher}{ACM}, \bibinfo{pages}{1–14}.
\newblock


\bibitem[Maddocks(2020)]%
        {Maddocks_2020}
\bibfield{author}{\bibinfo{person}{Sophie Maddocks}.} \bibinfo{year}{2020}\natexlab{}.
\newblock \showarticletitle{‘A Deepfake Porn Plot Intended to Silence Me’: exploring continuities between pornographic and ‘political’deep fakes}.
\newblock \bibinfo{journal}{\emph{Porn Studies}} \bibinfo{volume}{7}, \bibinfo{number}{4} (\bibinfo{year}{2020}), \bibinfo{pages}{415–423}.
\newblock


\bibitem[Madiega(2023)]%
        {Madiega_2023}
\bibfield{author}{\bibinfo{person}{Tambiama Madiega}.} \bibinfo{year}{2023}\natexlab{}.
\newblock \bibinfo{booktitle}{\emph{Artificial intelligence act}}.
\newblock


\bibitem[Malik and Malik(2021)]%
        {Malik_Malik_2021}
\bibfield{author}{\bibinfo{person}{Maya Malik} {and} \bibinfo{person}{Momin~M. Malik}.} \bibinfo{year}{2021}\natexlab{}.
\newblock \showarticletitle{Critical Technical Awakenings}.
\newblock \bibinfo{journal}{\emph{Journal of Social Computing}} \bibinfo{volume}{2}, \bibinfo{number}{4} (\bibinfo{date}{Dec.} \bibinfo{year}{2021}), \bibinfo{pages}{365–384}.
\newblock
\showISSN{2688-5255}
\urldef\tempurl%
\url{https://doi.org/10.23919/JSC.2021.0035}
\showDOI{\tempurl}


\bibitem[McGregor(2021)]%
        {McGregor_2021}
\bibfield{author}{\bibinfo{person}{Sean McGregor}.} \bibinfo{year}{2021}\natexlab{}.
\newblock \showarticletitle{Preventing Repeated Real World AI Failures by Cataloging Incidents: The AI Incident Database}.
\newblock \bibinfo{journal}{\emph{AAAI Conference on Artificial Intelligence}} \bibinfo{volume}{35}, \bibinfo{number}{1717} (\bibinfo{date}{May} \bibinfo{year}{2021}), \bibinfo{pages}{15458–15463}.
\newblock
\showISSN{2374-3468}
\urldef\tempurl%
\url{https://doi.org/10.1609/aaai.v35i17.17817}
\showDOI{\tempurl}


\bibitem[Metcalf et~al\mbox{.}(2019)]%
        {Metcalf_Moss_boyd_2019}
\bibfield{author}{\bibinfo{person}{Jacob Metcalf}, \bibinfo{person}{Emanuel Moss}, {and} \bibinfo{person}{danah boyd}.} \bibinfo{year}{2019}\natexlab{}.
\newblock \showarticletitle{Owning Ethics: Corporate Logics, Silicon Valley, and the Institutionalization of Ethics}.
\newblock \bibinfo{journal}{\emph{Social Research: An International Quarterly}} \bibinfo{volume}{86}, \bibinfo{number}{2} (\bibinfo{year}{2019}), \bibinfo{pages}{29}.
\newblock


\bibitem[Metz and Wakabayashi(2020)]%
        {Metz_Wakabayashi_2020}
\bibfield{author}{\bibinfo{person}{Cade Metz} {and} \bibinfo{person}{Daisuke Wakabayashi}.} \bibinfo{year}{2020}\natexlab{}.
\newblock \showarticletitle{Google Researcher Says She Was Fired Over Paper Highlighting Bias in A.I.}
\newblock \bibinfo{journal}{\emph{The New York Times}} (\bibinfo{date}{Dec.} \bibinfo{year}{2020}).
\newblock
\showISSN{0362-4331}
\urldef\tempurl%
\url{https://www.nytimes.com/2020/12/03/technology/google-researcher-timnit-gebru.html}
\showURL{%
\tempurl}


\bibitem[Mitchell et~al\mbox{.}(2019)]%
        {Mitchell_Wu_Zaldivar_Barnes_Vasserman_Hutchinson_Spitzer_Raji_Gebru_2019}
\bibfield{author}{\bibinfo{person}{Margaret Mitchell}, \bibinfo{person}{Simone Wu}, \bibinfo{person}{Andrew Zaldivar}, \bibinfo{person}{Parker Barnes}, \bibinfo{person}{Lucy Vasserman}, \bibinfo{person}{Ben Hutchinson}, \bibinfo{person}{Elena Spitzer}, \bibinfo{person}{Inioluwa~Deborah Raji}, {and} \bibinfo{person}{Timnit Gebru}.} \bibinfo{year}{2019}\natexlab{}.
\newblock \showarticletitle{Model Cards for Model Reporting}. In \bibinfo{booktitle}{\emph{Conference on Fairness, Accountability, and Transparency}}. \bibinfo{publisher}{ACM}, \bibinfo{address}{Atlanta GA USA}, \bibinfo{pages}{220–229}.
\newblock
\showISBNx{978-1-4503-6125-5}
\urldef\tempurl%
\url{https://doi.org/10.1145/3287560.3287596}
\showDOI{\tempurl}


\bibitem[Mittelstadt(2019)]%
        {mittelstadt2019principlesc}
\bibfield{author}{\bibinfo{person}{Brent Mittelstadt}.} \bibinfo{year}{2019}\natexlab{}.
\newblock \showarticletitle{Principles Alone Cannot Guarantee Ethical {{AI}}}.
\newblock \bibinfo{journal}{\emph{Nature Machine Intelligence}} \bibinfo{volume}{1}, \bibinfo{number}{11} (\bibinfo{date}{Nov.} \bibinfo{year}{2019}), \bibinfo{pages}{501--507}.
\newblock
\showISSN{2522-5839}
\urldef\tempurl%
\url{https://doi.org/10.1038/s42256-019-0114-4}
\showDOI{\tempurl}


\bibitem[Nafus(2012)]%
        {Nafus_2012}
\bibfield{author}{\bibinfo{person}{Dawn Nafus}.} \bibinfo{year}{2012}\natexlab{}.
\newblock \showarticletitle{‘Patches don’t have gender’: What is not open in open source software}.
\newblock \bibinfo{journal}{\emph{New Media \& Society}} \bibinfo{volume}{14}, \bibinfo{number}{4} (\bibinfo{year}{2012}), \bibinfo{pages}{669–683}.
\newblock


\bibitem[Nafus and Sherman(2014)]%
        {Nafus_Sherman_2014}
\bibfield{author}{\bibinfo{person}{Dawn Nafus} {and} \bibinfo{person}{Jamie Sherman}.} \bibinfo{year}{2014}\natexlab{}.
\newblock \showarticletitle{This One Does Not Go Up to 11: The Quantified Self Movement as an Alternative Big Data Practice}.
\newblock  (\bibinfo{year}{2014}), \bibinfo{pages}{11}.
\newblock


\bibitem[Narayanan(2022)]%
        {narayanan2022limits}
\bibfield{author}{\bibinfo{person}{Arvind Narayanan}.} \bibinfo{year}{2022}\natexlab{}.
\newblock \bibinfo{title}{The Limits of the Quantitative Approach to Discrimination}.  (\bibinfo{year}{2022}).
\newblock
\urldef\tempurl%
\url{https://www.cs.princeton.edu/~arvindn/talks/baldwin-discrimination/baldwin-discrimination-transcript.pdf}
\showURL{%
\tempurl}


\bibitem[NIST(2022)]%
        {NIST_2022}
\bibfield{author}{\bibinfo{person}{NIST}.} \bibinfo{year}{2022}\natexlab{}.
\newblock \showarticletitle{AI Risk Management Framework: Second Draft - August 18, 2022}.
\newblock  (\bibinfo{year}{2022}).
\newblock


\bibitem[Olson(2000)]%
        {Olson_2000}
\bibfield{author}{\bibinfo{person}{Lester~C. Olson}.} \bibinfo{year}{2000}\natexlab{}.
\newblock \showarticletitle{The Personal, the Political, and Others: Audre Lorde Denouncing “The Second Sex Conference”}.
\newblock \bibinfo{journal}{\emph{Philosophy \& Rhetoric}} \bibinfo{volume}{33}, \bibinfo{number}{3} (\bibinfo{year}{2000}), \bibinfo{pages}{259–285}.
\newblock
\showISSN{0031-8213}


\bibitem[Ottinger et~al\mbox{.}(2023)]%
        {Ottinger_Bronson_Nafus_2023}
\bibfield{author}{\bibinfo{person}{Gwen Ottinger}, \bibinfo{person}{Kelly Bronson}, {and} \bibinfo{person}{Dawn Nafus}.} \bibinfo{year}{2023}\natexlab{}.
\newblock \showarticletitle{Clicks and particulates: Value, alienation, and attunement as unifying themes in big data studies}.
\newblock \bibinfo{journal}{\emph{Big Data \& Society}} \bibinfo{volume}{10}, \bibinfo{number}{1} (\bibinfo{date}{Jan.} \bibinfo{year}{2023}), \bibinfo{pages}{20539517231184891}.
\newblock
\showISSN{2053-9517}
\urldef\tempurl%
\url{https://doi.org/10.1177/20539517231184891}
\showDOI{\tempurl}


\bibitem[Palinkas et~al\mbox{.}(2015)]%
        {palinkas2015purposeful}
\bibfield{author}{\bibinfo{person}{Lawrence~A. Palinkas}, \bibinfo{person}{Sarah~M. Horwitz}, \bibinfo{person}{Carla~A. Green}, \bibinfo{person}{Jennifer~P. Wisdom}, \bibinfo{person}{Naihua Duan}, {and} \bibinfo{person}{Kimberly Hoagwood}.} \bibinfo{year}{2015}\natexlab{}.
\newblock \showarticletitle{Purposeful {{Sampling}} for {{Qualitative Data Collection}} and {{Analysis}} in {{Mixed Method Implementation Research}}}.
\newblock \bibinfo{journal}{\emph{Administration and Policy in Mental Health and Mental Health Services Research}} \bibinfo{volume}{42}, \bibinfo{number}{5} (\bibinfo{date}{Sept.} \bibinfo{year}{2015}), \bibinfo{pages}{533--544}.
\newblock
\showISSN{1573-3289}
\urldef\tempurl%
\url{https://doi.org/10.1007/s10488-013-0528-y}
\showDOI{\tempurl}


\bibitem[Png(2022)]%
        {png2022tensions}
\bibfield{author}{\bibinfo{person}{Marie-Therese Png}.} \bibinfo{year}{2022}\natexlab{}.
\newblock \showarticletitle{At the {{Tensions}} of {{South}} and {{North}}: {{Critical Roles}} of {{Global South Stakeholders}} in {{AI Governance}}}. In \bibinfo{booktitle}{\emph{{{Conference}} on {{Fairness}}, {{Accountability}}, and {{Transparency}}}} \emph{(\bibinfo{series}{{{FAccT}} '22})}. \bibinfo{publisher}{{Association for Computing Machinery}}, \bibinfo{address}{{New York, NY, USA}}, \bibinfo{pages}{1434--1445}.
\newblock
\showISBNx{978-1-4503-9352-2}
\urldef\tempurl%
\url{https://doi.org/10.1145/3531146.3533200}
\showDOI{\tempurl}


\bibitem[Powdermaker(1966)]%
        {Powdermaker_1966}
\bibfield{author}{\bibinfo{person}{Hortense Powdermaker}.} \bibinfo{year}{1966}\natexlab{}.
\newblock \bibinfo{booktitle}{\emph{Stranger and Friend: The Way of an Anthropologist}}.
\newblock \bibinfo{publisher}{W. W. Norton \& Company}.
\newblock
\showISBNx{978-0-393-00410-6}


\bibitem[Pruss(2023)]%
        {pruss2023ghosting}
\bibfield{author}{\bibinfo{person}{Dasha Pruss}.} \bibinfo{year}{2023}\natexlab{}.
\newblock \showarticletitle{Ghosting the {{Machine}}: {{Judicial Resistance}} to a {{Recidivism Risk Assessment Instrument}}}. In \bibinfo{booktitle}{\emph{{{ACM Conference}} on {{Fairness}}, {{Accountability}}, and {{Transparency}}}} \emph{(\bibinfo{series}{{{FAccT}} '23})}. \bibinfo{publisher}{{Association for Computing Machinery}}, \bibinfo{address}{{New York, NY, USA}}, \bibinfo{pages}{312--323}.
\newblock
\showISBNx{9798400701924}
\urldef\tempurl%
\url{https://doi.org/10.1145/3593013.3593999}
\showDOI{\tempurl}


\bibitem[Raji et~al\mbox{.}(2022)]%
        {Raji_Kumar_Horowitz_Selbst_2022}
\bibfield{author}{\bibinfo{person}{Inioluwa~Deborah Raji}, \bibinfo{person}{I.~Elizabeth Kumar}, \bibinfo{person}{Aaron Horowitz}, {and} \bibinfo{person}{Andrew Selbst}.} \bibinfo{year}{2022}\natexlab{}.
\newblock \showarticletitle{The Fallacy of AI Functionality}. In \bibinfo{booktitle}{\emph{Conference on Fairness, Accountability, and Transparency}}. \bibinfo{publisher}{ACM}, \bibinfo{address}{Seoul Republic of Korea}, \bibinfo{pages}{959–972}.
\newblock
\showISBNx{978-1-4503-9352-2}
\urldef\tempurl%
\url{https://doi.org/10.1145/3531146.3533158}
\showDOI{\tempurl}


\bibitem[Raji et~al\mbox{.}(2021)]%
        {raji2021you}
\bibfield{author}{\bibinfo{person}{Inioluwa~Deborah Raji}, \bibinfo{person}{Morgan~Klaus Scheuerman}, {and} \bibinfo{person}{Razvan Amironesei}.} \bibinfo{year}{2021}\natexlab{}.
\newblock \showarticletitle{You {{Can}}'t {{Sit With Us}}: {{Exclusionary Pedagogy}} in {{AI Ethics Education}}}. In \bibinfo{booktitle}{\emph{{{Conference}} on {{Fairness}}, {{Accountability}}, and {{Transparency}}}}. \bibinfo{publisher}{{ACM}}, \bibinfo{address}{{Virtual Event Canada}}, \bibinfo{pages}{515--525}.
\newblock
\showISBNx{978-1-4503-8309-7}
\urldef\tempurl%
\url{https://doi.org/10.1145/3442188.3445914}
\showDOI{\tempurl}


\bibitem[Sambasivan et~al\mbox{.}(2021)]%
        {Sambasivan_Kapania_Highfill_Akrong_Paritosh_Aroyo_2021}
\bibfield{author}{\bibinfo{person}{Nithya Sambasivan}, \bibinfo{person}{Shivani Kapania}, \bibinfo{person}{Hannah Highfill}, \bibinfo{person}{Diana Akrong}, \bibinfo{person}{Praveen Paritosh}, {and} \bibinfo{person}{Lora~M Aroyo}.} \bibinfo{year}{2021}\natexlab{}.
\newblock \showarticletitle{“Everyone wants to do the model work, not the data work”: Data Cascades in High-Stakes AI}. In \bibinfo{booktitle}{\emph{Conference on Human Factors in Computing Systems}}. \bibinfo{publisher}{ACM}, \bibinfo{pages}{1–15}.
\newblock


\bibitem[Sampson et~al\mbox{.}(2023)]%
        {sampson2023representation}
\bibfield{author}{\bibinfo{person}{Princess Sampson}, \bibinfo{person}{Ro Encarnacion}, {and} \bibinfo{person}{Dana{\"e} Metaxa}.} \bibinfo{year}{2023}\natexlab{}.
\newblock \showarticletitle{Representation, {{Self-Determination}}, and {{Refusal}}: {{Queer People}}'s {{Experiences}} with {{Targeted Advertising}}}. In \bibinfo{booktitle}{\emph{2023 {{ACM Conference}} on {{Fairness}}, {{Accountability}}, and {{Transparency}}}}. \bibinfo{publisher}{{ACM}}, \bibinfo{address}{{Chicago IL USA}}, \bibinfo{pages}{1711--1722}.
\newblock
\showISBNx{9798400701924}
\urldef\tempurl%
\url{https://doi.org/10.1145/3593013.3594110}
\showDOI{\tempurl}


\bibitem[Schiffer and Newton(2023)]%
        {Schiffer_Newton_2023}
\bibfield{author}{\bibinfo{person}{Zoe Schiffer} {and} \bibinfo{person}{Casey Newton}.} \bibinfo{year}{2023}\natexlab{}.
\newblock \bibinfo{title}{Microsoft lays off team that taught employees how to make AI tools responsibly}.
\newblock
\newblock
\urldef\tempurl%
\url{https://www.theverge.com/2023/3/13/23638823/microsoft-ethics-society-team-responsible-ai-layoffs}
\showURL{%
\tempurl}


\bibitem[Schwandt(1994)]%
        {schwandt1994constructivist}
\bibfield{author}{\bibinfo{person}{Thomas Schwandt}.} \bibinfo{year}{1994}\natexlab{}.
\newblock \showarticletitle{Constructivist, {{Interpretivist Approaches}} to {{Human Inquiry}}}.
\newblock \bibinfo{journal}{\emph{Handbook of Qualitative Research Thousand Oaks, California: Sage}} (\bibinfo{year}{1994}).
\newblock


\bibitem[Seidman(2006)]%
        {seidman2006interviewing}
\bibfield{author}{\bibinfo{person}{Irving Seidman}.} \bibinfo{year}{2006}\natexlab{}.
\newblock \bibinfo{booktitle}{\emph{Interviewing as Qualitative Research: A Guide for Researchers in Education and the Social Sciences} (\bibinfo{edition}{3rd ed} ed.)}.
\newblock \bibinfo{publisher}{Teachers College Press}.
\newblock


\bibitem[Sloane et~al\mbox{.}(2022)]%
        {Sloane_Moss_Awomolo_Forlano_2022}
\bibfield{author}{\bibinfo{person}{Mona Sloane}, \bibinfo{person}{Emanuel Moss}, \bibinfo{person}{Olaitan Awomolo}, {and} \bibinfo{person}{Laura Forlano}.} \bibinfo{year}{2022}\natexlab{}.
\newblock \showarticletitle{Participation Is not a Design Fix for Machine Learning}. In \bibinfo{booktitle}{\emph{Equity and Access in Algorithms, Mechanisms, and Optimization}}. \bibinfo{publisher}{ACM}, \bibinfo{address}{Arlington VA USA}, \bibinfo{pages}{1–6}.
\newblock
\showISBNx{978-1-4503-9477-2}
\urldef\tempurl%
\url{https://doi.org/10.1145/3551624.3555285}
\showDOI{\tempurl}


\bibitem[Suchman(2002)]%
        {Suchman_2002}
\bibfield{author}{\bibinfo{person}{Lucy Suchman}.} \bibinfo{year}{2002}\natexlab{}.
\newblock \showarticletitle{Located accountabilities in technology production}.
\newblock \bibinfo{journal}{\emph{Scandinavian Journal of Information Systems}}  \bibinfo{volume}{14} (\bibinfo{year}{2002}), \bibinfo{pages}{16}.
\newblock
Issue 2.


\bibitem[Suchman(2023)]%
        {Suchman_2023}
\bibfield{author}{\bibinfo{person}{Lucy Suchman}.} \bibinfo{year}{2023}\natexlab{}.
\newblock \showarticletitle{The uncontroversial ‘thingness’ of AI}.
\newblock \bibinfo{journal}{\emph{Big Data \& Society}} \bibinfo{volume}{10}, \bibinfo{number}{2} (\bibinfo{date}{July} \bibinfo{year}{2023}).
\newblock
\showISSN{2053-9517, 2053-9517}
\urldef\tempurl%
\url{https://doi.org/10.1177/20539517231206794}
\showDOI{\tempurl}


\bibitem[Suresh et~al\mbox{.}(2022)]%
        {Suresh2022}
\bibfield{author}{\bibinfo{person}{Harini Suresh}, \bibinfo{person}{Rajiv Movva}, \bibinfo{person}{Amelia~Lee Dogan}, \bibinfo{person}{Rahul Bhargava}, \bibinfo{person}{Isadora Cruxen}, \bibinfo{person}{Angeles~Martinez Cuba}, \bibinfo{person}{Guilia Taurino}, \bibinfo{person}{Wonyoung So}, {and} \bibinfo{person}{Catherine D’Ignazio}.} \bibinfo{year}{2022}\natexlab{}.
\newblock \showarticletitle{Towards Intersectional Feminist and Participatory ML: A Case Study in Supporting Feminicide Counterdata Collection}. In \bibinfo{booktitle}{\emph{Conference on Fairness, Accountability, and Transparency}}. \bibinfo{publisher}{Association for Computing Machinery}, \bibinfo{address}{New York, NY, USA}, \bibinfo{pages}{667–678}.
\newblock
\showISBNx{978-1-4503-9352-2}
\urldef\tempurl%
\url{https://doi.org/10.1145/3531146.3533132}
\showDOI{\tempurl}


\bibitem[Team(2023)]%
        {Gemini_Team_2023}
\bibfield{author}{\bibinfo{person}{Gemini Team}.} \bibinfo{year}{2023}\natexlab{}.
\newblock \bibinfo{booktitle}{\emph{Gemini: A Family of Highly Capable Multimodal Models}}.
\newblock
\urldef\tempurl%
\url{https://storage.googleapis.com/deepmind-media/gemini/gemini_1_report.pdf}
\showURL{%
\tempurl}


\bibitem[Verran(2013)]%
        {Verran_2013}
\bibfield{author}{\bibinfo{person}{Helen Verran}.} \bibinfo{year}{2013}\natexlab{}.
\newblock \bibinfo{booktitle}{\emph{Engagements between disparate knowledge traditions: Toward doing difference generatively and in good faith}}.
\newblock \bibinfo{publisher}{HSRC Press}, \bibinfo{address}{Cape Town , South Africa}, \bibinfo{pages}{141–161}.
\newblock
\showISBNx{978-0-7969-2428-5}


\bibitem[West et~al\mbox{.}(2019)]%
        {West_Whittaker_Crawford_2019}
\bibfield{author}{\bibinfo{person}{Sarah~Myers West}, \bibinfo{person}{Meredith Whittaker}, {and} \bibinfo{person}{Kate Crawford}.} \bibinfo{year}{2019}\natexlab{}.
\newblock \bibinfo{booktitle}{\emph{Discriminating Systems: Gender, Race, and Power in AI}}.
\newblock \bibinfo{publisher}{AI Now Institute}.
\newblock
\urldef\tempurl%
\url{https://ainowinstitute.org/publication/discriminating-systems-gender-race-and-power-in-ai-2}
\showURL{%
\tempurl}


\bibitem[Whittaker(2023)]%
        {Whittaker_2023}
\bibfield{author}{\bibinfo{person}{Meredith Whittaker}.} \bibinfo{year}{2023}\natexlab{}.
\newblock \showarticletitle{Origin Stories: Plantations, Computers, and Industrial Control}.
\newblock \bibinfo{journal}{\emph{Logic(s) Magazine}} \bibinfo{number}{19-supa dupa skies (move slow and heal things)} (\bibinfo{date}{May} \bibinfo{year}{2023}).
\newblock
\urldef\tempurl%
\url{https://logicmag.io/supa-dupa-skies/origin-stories-plantations-computers-and-industrial-control/}
\showURL{%
\tempurl}


\bibitem[Widder et~al\mbox{.}(2021)]%
        {widder2021trust}
\bibfield{author}{\bibinfo{person}{David~Gray Widder}, \bibinfo{person}{Laura Dabbish}, \bibinfo{person}{James~D. Herbsleb}, \bibinfo{person}{Alexandra Holloway}, {and} \bibinfo{person}{Scott Davidoff}.} \bibinfo{year}{2021}\natexlab{}.
\newblock \showarticletitle{Trust in {{Collaborative Automation}} in {{High Stakes Software Engineering Work}}: {{A Case Study}} at {{NASA}}}. In \bibinfo{booktitle}{\emph{{{Conference}} on {{Human Factors}} in {{Computing Systems}}}}. \bibinfo{publisher}{ACM}, \bibinfo{pages}{1--13}.
\newblock
\urldef\tempurl%
\url{https://dl.acm.org/doi/10.1145/3411764.3445650}
\showURL{%
\tempurl}


\bibitem[Widder and Goues(2024)]%
        {widder2024whata}
\bibfield{author}{\bibinfo{person}{David~Gray Widder} {and} \bibinfo{person}{Claire~Le Goues}.} \bibinfo{year}{2024}\natexlab{}.
\newblock \bibinfo{title}{What Is a "Bug"? {{On}} Subjectivity, Epistemic Power, and Implications for Software Research}.
\newblock
\newblock
\urldef\tempurl%
\url{https://doi.org/10.48550/arXiv.2402.08165}
\showDOI{\tempurl}
\showeprint[arxiv]{2402.08165}~[cs]


\bibitem[Widder and Nafus(2023)]%
        {Widder_Nafus_2023}
\bibfield{author}{\bibinfo{person}{David~Gray Widder} {and} \bibinfo{person}{Dawn Nafus}.} \bibinfo{year}{2023}\natexlab{}.
\newblock \showarticletitle{Dislocated accountabilities in the “AI supply chain”: Modularity and developers’ notions of responsibility}.
\newblock \bibinfo{journal}{\emph{SAGE Big Data \& Society}} \bibinfo{volume}{10}, \bibinfo{number}{1} (\bibinfo{date}{Jan.} \bibinfo{year}{2023}), \bibinfo{pages}{20539517231177620}.
\newblock
\showISSN{2053-9517, 2053-9517}
\urldef\tempurl%
\url{https://doi.org/10.1177/20539517231177620}
\showDOI{\tempurl}


\bibitem[Widder et~al\mbox{.}(2023)]%
        {Widder_Zhen_Dabbish_Herbsleb_2023}
\bibfield{author}{\bibinfo{person}{David~Gray Widder}, \bibinfo{person}{Derrick Zhen}, \bibinfo{person}{Laura Dabbish}, {and} \bibinfo{person}{James Herbsleb}.} \bibinfo{year}{2023}\natexlab{}.
\newblock \showarticletitle{It’s about power: What ethical concerns do software engineers have, and what do they (feel they can) do about them?}. In \bibinfo{booktitle}{\emph{Conference on Fairness, Accountability, and Transparency}}. \bibinfo{publisher}{ACM}, \bibinfo{address}{Chicago IL USA}, \bibinfo{pages}{467–479}.
\newblock
\showISBNx{9798400701924}
\urldef\tempurl%
\url{https://doi.org/10.1145/3593013.3594012}
\showDOI{\tempurl}


\bibitem[Williams et~al\mbox{.}(2014)]%
        {williams2014multisited}
\bibfield{author}{\bibinfo{person}{Amanda Williams}, \bibinfo{person}{Silvia Lindtner}, \bibinfo{person}{Ken Anderson}, {and} \bibinfo{person}{Paul Dourish}.} \bibinfo{year}{2014}\natexlab{}.
\newblock \showarticletitle{Multisited {{Design}}: {{An Analytical Lens}} for {{Transnational HCI}}}.
\newblock \bibinfo{journal}{\emph{Human{\textendash}Computer Interaction}} \bibinfo{volume}{29}, \bibinfo{number}{1} (\bibinfo{date}{Jan.} \bibinfo{year}{2014}), \bibinfo{pages}{78--108}.
\newblock
\showISSN{0737-0024, 1532-7051}
\urldef\tempurl%
\url{https://doi.org/10.1080/07370024.2013.823819}
\showDOI{\tempurl}


\bibitem[Yi(2021)]%
        {Yi_2021}
\bibfield{author}{\bibinfo{person}{Michelle Yi}.} \bibinfo{year}{2021}\natexlab{}.
\newblock \bibinfo{title}{Model Cards: A Nutrition Label for Predictive Models}.
\newblock
\newblock
\urldef\tempurl%
\url{https://www.michelleyi.ai/blog/model-cards-a-nutrition-label-for-predictive-models}
\showURL{%
\tempurl}


\bibitem[Young et~al\mbox{.}(2022)]%
        {Young_Katell_Krafft_2022}
\bibfield{author}{\bibinfo{person}{Meg Young}, \bibinfo{person}{Michael Katell}, {and} \bibinfo{person}{P.M. Krafft}.} \bibinfo{year}{2022}\natexlab{}.
\newblock \showarticletitle{Confronting Power and Corporate Capture at the FAccT Conference}. In \bibinfo{booktitle}{\emph{Conference on Fairness, Accountability, and Transparency}}. \bibinfo{publisher}{ACM}, \bibinfo{address}{Seoul Republic of Korea}, \bibinfo{pages}{1375–1386}.
\newblock
\showISBNx{978-1-4503-9352-2}
\urldef\tempurl%
\url{https://doi.org/10.1145/3531146.3533194}
\showDOI{\tempurl}


\end{thebibliography}

\appendix

\end{document}